\documentclass[fleqn,usenatbib]{mnras}
\usepackage{fix-cm}

\usepackage{newtxtext,newtxmath}

\usepackage[T1]{fontenc}
\usepackage{cancel}

\DeclareRobustCommand{\VAN}[3]{#2}
\let\VANthebibliography\thebibliography
\def\thebibliography{\DeclareRobustCommand{\VAN}[3]{##3}\VANthebibliography}

\usepackage{graphicx}	
\usepackage{amsmath}	

\newcommand{\ltsimeq}{\raisebox{-0.6ex}{$\,\stackrel
        {\raisebox{-.2ex}{$\textstyle <$}}{\sim}\,$}}
\newcommand{\gtsimeq}{\raisebox{-0.6ex}{$\,\stackrel
        {\raisebox{-.2ex}{$\textstyle >$}}{\sim}\,$}}

\title[No excess AGN activity in massive PSBs at cosmic noon]{No evidence for excess AGN activity in recently quenched massive galaxies at cosmic noon}

\author[O. Almaini, et al.]{Omar Almaini$^{1}$\thanks{Contact e-mail: \href{mailto:omar.almaini@nottingham.ac.uk}{omar.almaini@nottingham.ac.uk}},
Vivienne Wild$^{2}$,
David Maltby$^{1}$,
Elizabeth Taylor$^{1}$,
Kate Rowlands$^{3,4}$,\newauthor
Thomas de Lisle$^{1}$,
Katherine Alatalo$^{5,3}$,
Jimi Harrold$^{1}$,
Guillaume Hewitt$^{1}$, 
Pallavi Patil$^{3}$,\newauthor
Maya Skarbinski$^{3}$
\\
$^{1}$ School of Physics and Astronomy, University of Nottingham, University Park, Nottingham NG7 2RD, U.K.\\
$^{2}$ School of Physics and Astronomy, University of St Andrews, North Haugh, St Andrews, KY16 9SS, U.K.\\
$^{3}$ William H. Miller III Department of Physics and Astronomy, Johns Hopkins University,  Baltimore, MD 21218, USA \\
$^{4}$ AURA for ESA, Space Telescope Science Institute, 3700 San Martin Dr., Baltimore, MD 21218, USA \\
$^{5}$ Space Telescope Science Institute, 3700 San Martin Dr., Baltimore, MD 21218, USA 
}

\date{Accepted XXX. Received YYY; in original form ZZZ}

\pubyear{\the\year{}}

\begin{document}
\label{firstpage}
\pagerange{\pageref{firstpage}--\pageref{lastpage}}
\maketitle

\begin{abstract}
We present an analysis of AGN activity within recently quenched massive galaxies at cosmic noon ($z\sim 2$),  using deep \textit{Chandra} X-ray observations of the Ultra-Deep Survey (UDS) field. Our sample includes over 4000 massive galaxies ($M_\ast > 10^{10.5}$ M$_{\odot}$) in the redshift range $1<z<3$, including more than 200 transitionary post-starburst (PSB) systems. We find that X-ray emitting AGN are detected in  $6.2 \pm 1.5$\% of massive PSBs at these redshifts, a detection rate that lies between those of star-forming and passive galaxies ($8.2 \pm 0.5$\% and $5.7 \pm 0.8$\%,  respectively). A stacking analysis shows that the average X-ray luminosity for PSBs is comparable to older passive galaxies, but a factor of $2.6\pm 0.3$ below star-forming galaxies of similar redshift and stellar mass. 
The average X-ray luminosity in all populations appears to trace the star-formation rate, with PSBs showing low levels of AGN activity consistent with their reduced levels of star formation.
We conclude that, on average, we see no evidence for excess AGN activity in the post-starburst phase. However, the low levels of AGN activity can be reconciled with the high-velocity
outflows observed in many PSBs, assuming the rare X-ray detections represent short-lived bursts of black hole activity, visible 
$\sim$5\% of the time. Thus, X-ray AGN may help to maintain quiescence in massive galaxies at cosmic noon, but the evidence for a direct link to the primary quenching event remains elusive.
\end{abstract}




\section{Introduction}

A striking feature of galaxies in the local Universe is their strong bimodality. Galaxies at high stellar mass
($M_\ast>10^{10.5}$ M$_{\odot}$) are typically spheroidal in shape, with old stellar populations,  while lower-mass galaxies are mostly disc-like, with younger stellar populations \citep[e.g.,][]{Strateva2001, Baldry2004}. Deep infrared surveys have allowed us to probe the build-up of these galaxy populations over most of the history of the Universe \citep[e.g.,][]{Fontana2004, Ilbert2013, McLeod2021}, revealing that most local massive galaxies terminated their star formation at early times, during an epoch commonly referred to as `cosmic noon' ($1<z<3$). 

Active galactic nucleus (AGN) feedback is widely believed to be responsible for quenching star formation in massive galaxies, and indeed over its lifetime a growing supermassive black hole will radiate many times the binding energy of its host galaxy \citep[e.g.,][]{Silk1998}. Models of AGN feedback typically invoke a high-accretion mode ("quasar-mode") to quench star formation at high redshift, followed by a "radio-mode" at lower redshifts to prevent the accretion of further cold gas \citep[e.g.,][]{Croton2006, Hopkins2006, Bower2006, Dave2019}, but the precise mechanisms driving the initial quenching at cosmic noon are not well understood.

Post-starburst galaxies (PSBs) are potentially an ideal population for 
testing our understanding of the quenching process, as they are believed to represent galaxies in transition between the star-forming and passive populations. PSBs have spectral energy distributions (SEDs) dominated by A/F-type stars, with little ongoing star formation, indicating a system in which a major burst of star formation was rapidly quenched within the last Gyr \citep[e.g.,][]{Dressler1983, Wild2020, French2021}.

High-redshift ($z>1$) PSBs can now be identified in large numbers using photometric methods, given sufficiently deep multi-band imaging
either side of the Balmer-break region. Their distinctive A-star dominated 
SEDs (which appear "triangular" in $\log f_{\lambda}$) allow PSBs to be reliably separated from star-forming galaxies and older passive galaxies, e.g., by using 
principal component analysis (PCA) techniques \citep[][]{Wild2014, Maltby2016, Wild2020}. Studies of the resulting stellar mass functions  \citep[][]{Wild2016,Taylor2023}, large-scale clustering \citep[][]{Wilkinson2021}, and star-formation histories \citep[][]{Belli2019, Wild2020}, suggest that massive ($M_\ast>10^{10.5}$ M$_{\odot}$) PSBs are consistent with being the immediate descendants of luminous submillimetre galaxies
\citep[SMGs, e.g.,][]{Smail97,Hughes98, Swinbank2014, Dudzevicute2020}, having undergone an intense period of starburst activity prior to a rapid quenching event.

Studies of the morphologies of high-z PSBs provide further clues to their origin. At $z>1$, 
massive  ($M_\ast>10^{10.5}$ M$_{\odot}$) PSBs 
 are found to be exceptionally compact and spheroidal, strongly suggesting  they have formed from the dissipative collapse of gas, e.g., following a gas-rich merger
\citep[][]{Whitaker2012, Almaini2017, Maltby2018}.
For a gas-rich merger, SPH simulations predict a distinct delay of $\sim$100 Myr (or longer) between the peak of star formation and the peak in AGN activity \citep{Hopkins2012}, 
suggesting that high-z PSBs should be ideal locations for identifying the signatures of AGN feedback.
In the local Universe there is indeed evidence that AGN activity peaks a few hundred Myr after the peak in star formation 
\citep[e.g.,][]{Davies2007, Wild2010}.

Among local PSBs, many studies have found evidence for LINER-like activity \citep[e.g.,][]{Yan2006, Wild2007}, but it is often uncertain whether this is due to AGN or other phenomena, such as shocks or hot evolved low-mass stars \citep[e.g.,][]{Stasinska2008, CidFernandes2011, Belfiore2016}.
X-ray studies of low-redshift PSBs have revealed evidence for weak AGN 
in up to one third of these galaxies \citep[][]{Georgakakis2008, Brown2009, Lanz2022}, 
but this low-level activity may indicate AGN that are "along for the ride" \citep[to quote][]{Lanz2022}, and simply tracing the reduced availability of gas.  Using optical line ratio diagnostics, \cite{Yesuf2014} find evidence for AGN in around one third of local PSBs, but conclude that the AGN are not responsible for the original quenching of star formation.
\cite{French2023} conducted a study of Extended Emission Line Regions (EELRs) in local PSBs to estimate the duty cycle of luminous AGN activity, concluding that the central AGN must have very short duty cycles, and are only "on" approximately 5 per cent of the time. They argue that such short duty cycles may explain the apparently low level of AGN activity observed in local PSBs. However, some studies with different selection methods suggest much longer duty cycles, with AGN detected in up to 50\% of PSBs \citep{Sell2014,Pawlik2018}.

At higher redshift, the launch of  
\textit{JWST} has allowed the first detailed spectroscopic studies of quenched galaxies at $z>2$. So far a number of recent studies have reported evidence for a high AGN fraction in these galaxies,  based on line ratio diagnostics \citep[e.g.,][]{Belli2024, Davies2024, D'Eugenio2024, Wu2024}. 
The samples are relatively small so far, however, 
so it is unclear if these results represent 
direct evidence in favour of the AGN feedback scenario. Massive high-z PSBs also show evidence for high-velocity outflows 
\citep[e.g.,][]{Tremonti2007, Maltby2019, Belli2024, Davies2024, D'Eugenio2024},
which strongly suggest AGN-driven winds.
However, these winds 
may be caused by later episodes of AGN activity, rather than necessarily being linked to the primary quenching event. The recent discovery of high-velocity outflows in relatively `old' PSBs ($>500$\, Myr
after the burst) would indicate that strong winds can be driven long after quenching has ended \citep{Taylor2024}. 

In this work, we undertake the first systematic study of X-ray AGN activity in a large sample of PSBs at cosmic noon, identified 
from the UKIDSS Ultra-Deep Survey \citep[][]{Lawrence2007, Almaini2017} using the photometric PCA analysis described in \citet{Wild2014} and \cite{Wilkinson2021}. We use deep \textit{Chandra} observations to measure AGN activity, 
both from individual detections and from a stacking analysis. 
We compare the results to large control samples of star-forming and passive galaxies of equivalent redshift and stellar mass, to determine whether there is any evidence for excess AGN acitivity associated with recent quenching.

Throughout this paper, we adopt a flat $\Lambda$CDM cosmology with $\Omega_M$ = 0.3, $\Omega_{\Lambda}$ = 0.7, and $H_0 = 70$\,kms$^{-1}$ Mpc$^{-1}$.

\section{Data and sample selection}

\subsection{K-band galaxy catalogue}
\label{uds}
\label{subsec:dr11}

We use a $K$-band selected galaxy sample from the UKIDSS Ultra-Deep Survey  \protect\citep[UDS;][]{Lawrence2007}. The 11th UDS data release (DR11) reached $5\sigma$ near-infrared depths of $J= 25.6$, $H= 25.1$, and $K = 25.3$ over $0.77$ sq degrees. To date, the UDS is the deepest contiguous $K$-band survey over such a large area, with a wide array of complementary multi-wavelength imaging. For the purposes of our study we use the $12$-band photometric catalogue described in \cite{Wilkinson2021}, which combines UDS imaging with optical imaging from the Subaru telescope in $B, V, R ,I’, z’$
\citep[][]{Furusawa2008}, $u$-band imaging from CFHT, $Y$-band imaging from VISTA 
\citep[][]{Jarvis2013}, and \textit{Spitzer} IRAC imaging at 3.6 and 4.5$\mu$m from the SpUDS Legacy Program, combined with deeper IRAC imaging
from the Spitzer Extended Deep Survey \citep[SEDS;][]{Ashby2013}.
Allowing for the masking of bright stars and artefacts, the overlapping area with $12$-band photometry is $0.63$ sq degrees.

The method used to determine photometric redshifts is described in \cite{Wilkinson2021}. Briefly, the EAzY code \citep[][]{Brammer2008} was used with the default template configuration of 12 Flexible Stellar Population Synthesis (FSPS) SED components 
\citep[][]{Conroy2010}, with the addition of three simple stellar population (SSP) templates.
The three SSP models have ages of 20, 50 and 150 Myr, using a Chabrier IMF and sub-solar metallicity (0.2 solar), to
represent recent bursts of star formation to complement continuous star-formation histories in the FSPS templates. Photometric redshifts were calibrated using approximately 8000 sources with secure spectroscopic redshifts, and comparison with the photometric redshifts yields a dispersion in ($z_{\rm{phot}} - z_{\rm {spec}})/(1 + z_{\rm {spec}}$) of $\sigma_{\rm{NMAD}} = 0.019$. Further details can be found in \cite{Wilkinson2021} and \cite{Taylor2023}.

\subsection{Galaxy properties and classifications}
\label{subsec:galaxies}

We use the first three ``super-colours'' from \citep{Wilkinson2021} to distinguish galaxies with different recent star formation histories. ``Super-colours'' are linear combinations of the broad-band photometry fluxes, with weighting vectors built from a Principal Component Analysis (PCA) to reduce the dimensionality of the multi-band photometric dataset while maximising retention of variance. Unlike traditional rest-frame colour-colour diagnostic plots (e.g., UVJ), super-colours make full use of all bands to maximise the separation of populations with distinct spectral energy distribution (SED) shapes, and do not force data to fit imperfect spectral synthesis models. Originally created for UDS-DR8 \citep{Wild2014,Wild2016}, we use the updated DR11 catalogue from \citep{Wilkinson2021} which extends the rest-frame wavelength range of filters used to 2500-15000\AA, allowing robust rest-frame spectral energy distribution (SED) characterisation over the redshift range $0.5<z<3$. Redshifts are taken from spectroscopy where available, or otherwise from  the photometric redshift catalogue described in Section \ref{subsec:dr11}. We use the first two super-colours to classify quenched, star-forming and post-starburst galaxies. The class of ``dusty'' galaxies identified by \citet{Wild2014} was combined into the star-forming class for the purposes of this analysis. 

As described in \citet{Wild2014} we fit the first three super-colours to a large grid of \citet{Bruzual2003} spectral synthesis models covering a wide range of bursty star formation histories, metallicities, and \citet{Charlot2000} two-component dust attenuation. The resulting probability distribution functions (PDFs) of derived (model-dependent) properties provide median and errors (16th and 84th percentile) of physical properties such as stellar mass, star-formation rate (averaged over 100\,Myr), time since the last starburst,  mean light-weighted age, and dust attenuation.

\subsection{Chandra X-ray data and counterparts}
\label{section:chandra}
To identify X-ray emitting AGN we use \textit{Chandra} data from the X-UDS survey
\citep[PI: G. Hasinger;][]{Kocevski2018}. The survey consists of 25 Chandra/ACIS-I observations, each of approximately 50ks, covering a total area of $0.33$\,deg$^2$. The imaging mosaic was designed to achieve an exposure time of $\sim 600$\,ks in the innermost region  ($\sim 0.06$\,deg$^2$), and  $\sim 200$\,ks  over the majority of the remainder of the field. In total, 868 X-ray sources were identified \citep{Kocevski2018}.

We match the \textit{Chandra} X-ray catalogue of \citet[][]{Kocevski2018} to the UDS DR11 $K$-band catalogue 
 \citep{Wilkinson2021} using the maximum likelihood method described in 
 \cite{Sutherland1992}.
 For every X-ray source and possible $K$-band counterpart, this method takes into account both the separation, the space density of background galaxies at that magnitude, and a prior on the expected magnitude distribution of X-ray sources. The magnitude prior is determined from the distribution of $K$-band magnitudes obtained from the most secure matches obtained within  1-arcsec, after ensuring the \textit{Chandra} and UKIDSS astrometric frames are aligned. The resulting analysis produces a likelihood ratio (LR)  for every potential counterpart, defined as the ratio between the likelihood that the galaxy counterpart is the correct identification and the corresponding probability of selecting an unrelated background object. 

A likelihood threshold is then selected, chosen to maximise the overall completeness (C) while minimising the number of spurious identifications (by maximising a reliability parameter, R).  
Following previous authors 
\citep[e.g.,][]{Brusa2005,Civano2012},  we selected the likelihood threshold that maximised the sum of the reliability (R) and completeness (C) parameters, which gave a threshold $L_{th}>0.8$. 

We found that 734/868 X-ray sources had at least one $K$-band counterpart within 5 arcsec above the chosen likelihood threshold (of which 96\% were identified within 1.5 arcsec). Of these, 672 were unique matches, while 62 had two or more possible counterparts above the matching threshold. For the 62 with multiple matches, a secure identification was defined in 41 cases where the likelihood ratio exceeded all other potential matches by at least a factor of three. For the remaining 21 sources, a further 16 “probable” matches were assigned, for which the most likely $K$-band counterpart has a greater than 50\% probability of being the correct match. For the purposes of this work, we combine the resulting 713 secure and 16 probable matches to define an initial catalogue of 729 \textit{Chandra} X-ray sources with $K$-band identifications. 

The sensitivity of \textit{Chandra} is a strong function of off-axis angle, so we restrict our analysis to X-ray sources detected within 8 arcmin from one of the \textit{Chandra} field centres in the X-UDS mosaic.
This  radius was chosen to match 
the default region used by the CSTACK stacking tool \citep[][see Section \ref{section:stack}]{Miyaji2008},
to ensure consistency in our analysis. This restriction limits the analysis to the central 0.26 sq degrees of the X-UDS \textit{Chandra} mosaic, within which the median X-ray flux limit is  $8.4\times10^{-16} $\,erg s$^{-1}$\,cm$^{-2}$ ($0.5-8$\, keV), i.e. the flux limit is deeper than this value over half of the survey area \citep[see][]{Kocevski2018}.
There are 796 X-ray sources within this deeper region, of which 673 have $K$-band identifications.
We investigated restricting the analysis to progressively deeper (and smaller) regions based on the exposure map, but none of the results presented in this paper were significantly affected.

 Of the 673 X-ray sources with $K$-band identifications within the deep \textit{Chandra} region, 502 lie within the redshift range $0.5<z<3$ and have PCA galaxy classifications from 
 \cite{Wilkinson2021}.
  These classifications were determined for all galaxies within the redshift range $0.5<z<3$ to a $K$-band magnitude limit of $K_{AB}<24.5$, after excluding masked regions in the UDS imaging
  \citep[see][]{Wilkinson2021}. The final sample of 502 X-ray sources are associated with 385 star-forming galaxies, 83 passive galaxies, 25  PSBs, and 9 "unclassified" systems. The unclassified sample are galaxies for which the PCA analysis was unable to assign a reliable galaxy classification, due to colours lying outside the expected range for a galaxy SED.  
An inspection of these 9 galaxies reveals that 5 show very blue SEDs and point-like profiles, consistent with unresolved Type 1 quasars. We note that 2 are classified as bright quasars in the analysis presented in Section \ref{section:quasars}. Another two have very nearby companions, which are likely to be interlopers affecting the galaxy photometry. The 9 unclassified galaxies (1.8 per cent of the sample) are removed from our analysis, due to the uncertain nature of their host galaxies.

X-ray luminosities are determined for all X-ray  sources in the 
$0.5-8$\,keV band, $K$-corrected assuming a power-law photon index with a slope $\Gamma=1.7$.

\subsection{Removal of bright quasars}
\label{section:quasars}

The presence of AGN light can affect the determination of host galaxy properties, particularly for the most luminous Type 1 AGN. 
In a \textit{Chandra} X-ray survey of similar depth, \cite{Pierce2010} conclude that optical galaxy colours can be strongly affected by the presence of an AGN in up to 10\% of cases, typically for the most luminous, unobscured AGN that appear as point-like quasars.

A detailed decomposition of galaxy and AGN light is beyond the scope of this paper, though it may be possible in future work for the subset of galaxies now observed with deep multi-band \textit{JWST} imaging. As a pragmatic solution, we follow a similar approach to previous authors
\citep[e.g.,][]{Aird2012,Kocevski2017}, and choose to remove the most luminous, point-like AGN from our sample, for which galaxy classifications and stellar masses may be unreliable. We explore this issue in more detail in Section \ref{section:sims},  where we find that AGN can have the most significant impact on galaxy classification when the Eddington ratio exceeds 10\%, if the AGN has relatively low levels of nuclear reddening
($A_v \ltsimeq 1$). For a typical PSB in our sample, this Eddington ratio corresponds to an X-ray luminosity
$L_{0.5-8\, \rm{keV}}\simeq 10^{44}$ erg s$^{-1}$.
 We therefore choose (a priori) to remove all X-ray AGN above this X-ray luminosity that also display evidence for point-like profiles in our 
 $K$-band imaging (\texttt{CLASS\_STAR}\,$>0.95$), 
 to remove the most luminous "quasar-like" objects. 
 These criteria remove 35 out of 502 X-ray sources from our analysis,  of which 32 out of 407 ($\sim$ 8\%)
  are of high stellar mass ($M_\ast>10^{10.5}$ M$_{\odot}$), as used in the bulk of the later analysis. Formally, 26 of these galaxies were classified as star-forming galaxies in the PCA analysis, 2 are passive galaxies, 2 are PSBs, and 2 are unclassified.

We find, however, that our primary results do not depend strongly on the precise criteria used to remove these quasars. In Appendix \ref{section:appendix} we explore the impact of stricter and more relaxed criteria, finding that our primary conclusions are unaffected. Further discussion of the potentially misclassified galaxies can be found in Section \ref{section:sims}.

\section{Individual X-ray detections}
\label{section:detections}

\begin{figure}
 \includegraphics[width=\columnwidth]{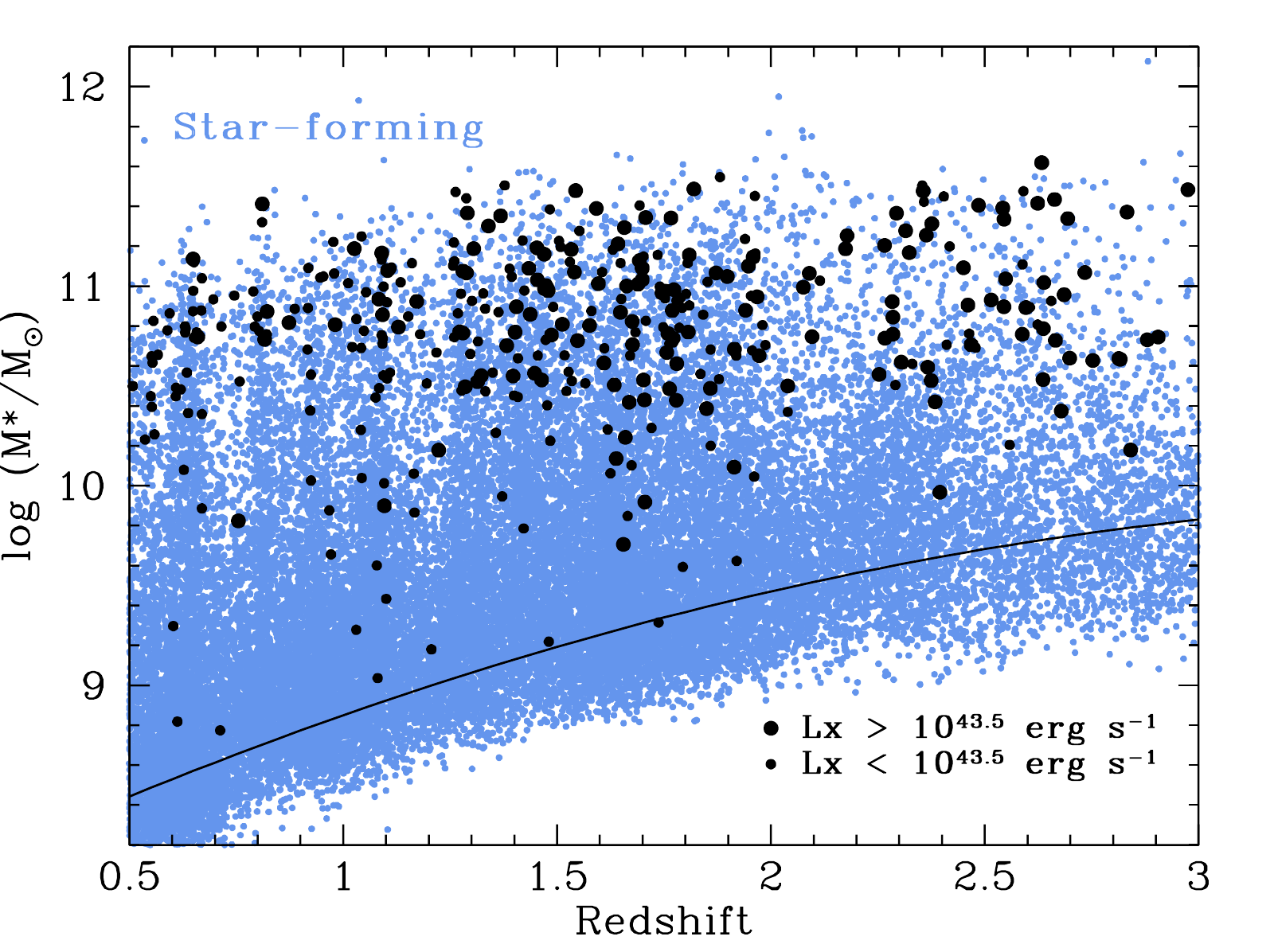}
  \includegraphics[width=\columnwidth]
 {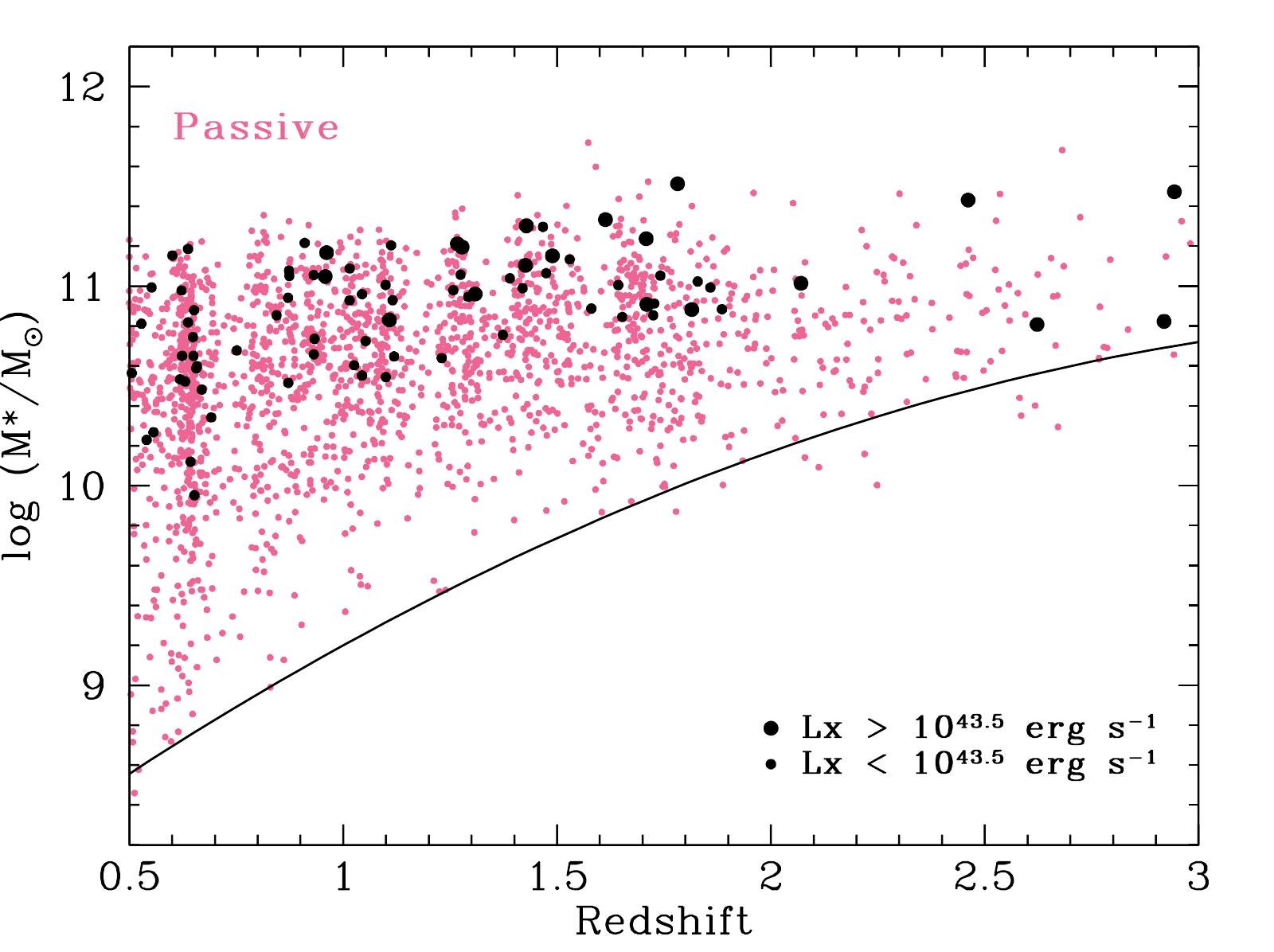}
  \includegraphics[width=\columnwidth]
 {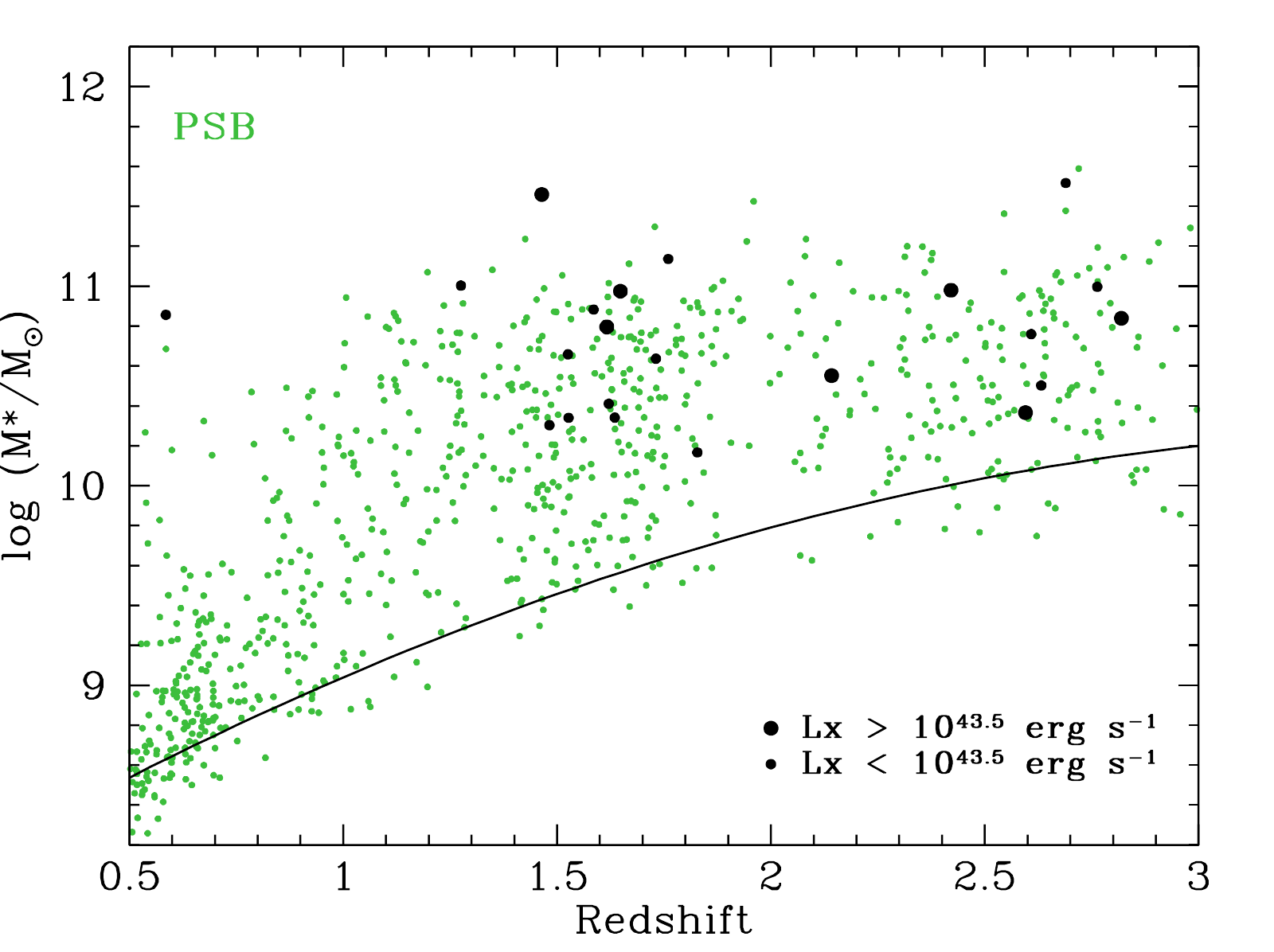}
 \caption{The stellar mass versus redshift relation for star-forming, passive, and post-starburst galaxies within the region of the UDS field with deep \textit{Chandra} X-ray imaging. 
 The solid black curves represent the 90\% stellar mass completeness limits, determined separately for each galaxy class. \textit{Chandra} X-ray detections are highlighted with solid black symbols, in two luminosity bins ($0.5-8$\,keV). The upper luminosity range ($L_{\rm X}> 10^{43.5}$ erg s$^{-1}$) corresponds to the approximate X-ray detection limit at $z=3$. 
 }
 \label{fig:zmass}
\end{figure}

In Figure \ref{fig:zmass} we present the stellar mass versus redshift relations for the 
star-forming, passive, and post-starburst galaxies, with \textit{Chandra} X-ray detections highlighted. Only galaxies located within the inner regions of the
\textit{Chandra} mosaic are used in this comparison, as described in Section \ref{section:chandra}.
 The small fraction of quasar-like AGN are removed, as outlined in Section \ref{section:quasars}, as these AGN may have unreliable stellar masses and galaxy spectral classifications.
The 90\% stellar mass completeness limits are illustrated (solid curves), determined using the method of \cite{Pozzetti2010}, as described in \cite{Wilkinson2021}.
These curves demonstrate that our survey is largely complete for all galaxy types above a stellar mass limit of 10$^{10.5}$\,M$_{\odot}$ to $z\sim3$. 
In order to study the prevalence of AGN as a function of redshift we therefore focus on galaxies above this stellar mass limit for the purposes of this study. 
Within the inner \textit{Chandra} region, these criteria leave a sample of $3522$ star-forming galaxies, $1296$ passive galaxies, and $245$ PSBs in the redshift range $0.5<z<3.0$, above a stellar mass of 10$^{10.5}$\,M$_{\odot}$ (of which 2916, 828, and 243 respectively lie at $z>1$).
It is worth noting that there are only 2 massive PSBs in our sample in the redshift range $0.5<z<1$, reflecting the very strong redshift evolution in this population 
\citep{Wild2016}.

\begin{figure}
 \includegraphics[width=\columnwidth]{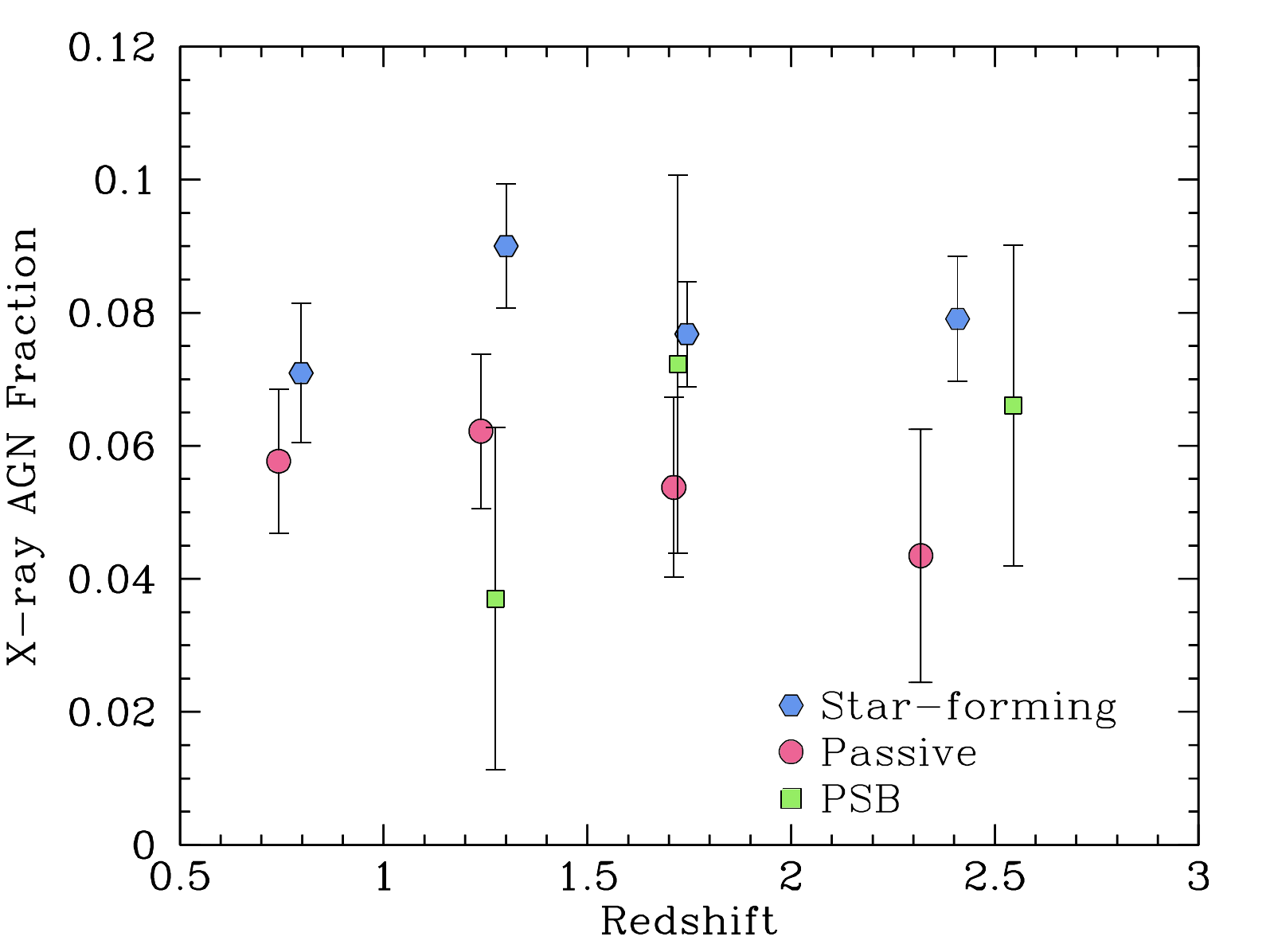}
  \includegraphics[width=\columnwidth]{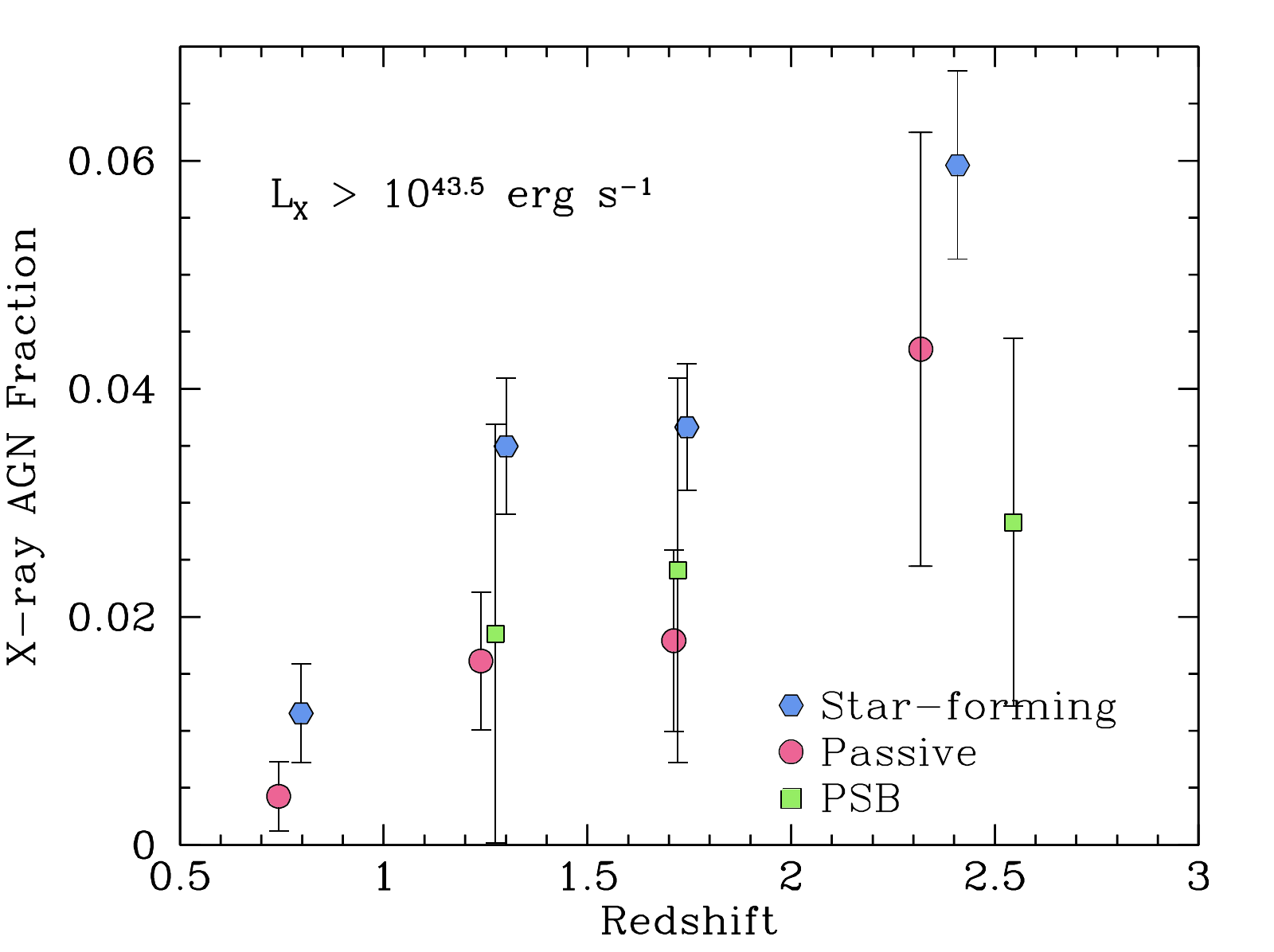}
 \caption{The X-ray detected fraction as a function of redshift for the three primary galaxy populations, for galaxies with high stellar mass ($M_\ast>10^{10.5}$ M$_{\odot}$). 
 Redshift bins of width $\Delta z=0.5$ are used in the range $0.5<z<2$, 
 with a single bin for the range $2<z<3$.
 Points are plotted at the mean redshift within each bin. 
 The upper plot is based on all X-ray detections, while the lower plot is restricted to  AGN with luminosities $L_{\rm X}> 10^{43.5}$ erg s$^{-1}$ (0.5-8 keV). 
This X-ray luminosity corresponds to the approximate detection limit at $z=3$, 
allowing a fairer comparison of X-ray activity between redshift bins. Note the difference in  scale on the two figures. All bins displayed contain a minimum of 50 galaxies. For clarity, the low-redshift PSB bin  containing only two galaxies is excluded. One of these PSBs is detected by \textit{Chandra} at low luminosity (see Figure \ref{fig:zmass}).}
\label{fig:frac}
\end{figure}

Figure \ref{fig:zmass}  confirms the well-known result that detected X-ray AGN are preferentially located in galaxies of high stellar mass 
    \citep[e.g.,][]{Alonso-Herrero2008, Aird2012}.
It is also clear that the majority of X-ray selected AGN at these redshifts are located
within star-forming galaxies.
In Figure \ref{fig:frac}  we explore the {\it fraction} of AGN as a function of redshift for each galaxy type, restricting the comparison to the most massive galaxies ($M_\ast>10^{10.5}$ M$_{\odot}$). The overall X-ray detection fraction is presented, along with the equivalent for detections at luminosities
$L_{\rm X}> 10^{43.5}$ erg s$^{-1}$ ($0.5-8$\, keV). 
This X-ray luminosity corresponds to the approximate \textit{Chandra}
detection limit at $z\sim 3$,  allowing a fairer comparison between redshift bins. 
Above this luminosity we find that the fraction of luminous AGN increases with redshift for all galaxy populations
(Figure \ref{fig:frac}, lower panel).
\footnote{Strictly speaking, this luminosity limit at $z=3$ is only reached over the deepest $\sim$30\% of the survey area used in our analysis. Therefore the highest redshift bin in the lower panel of Figure \ref{fig:frac} may be slightly incomplete, although the cross-comparison of the populations will still be valid. Correcting for this effect would boost the detected AGN fractions with $L_{\rm X}> 10^{43.5}$ at $z>2$ by an estimated
factor of $1.2$, i.e., within current uncertainties. }

Overall  we find that the fraction of galaxies hosting AGN is slightly higher 
in star-forming galaxies compared to passive galaxies at all redshifts, confirming previous results 
\citep[e.g.,][]{Silverman2009, Aird2019, Ni2023}.
For example, in the redshift range $1<z<3$ (`cosmic noon'), X-ray emitting AGN are detected in 8.2 $\pm$ 0.5\%
of star-forming galaxies and  5.7 $\pm$ 0.8\%  of passive galaxies. The AGN fraction in post-starburst galaxies is 6.2 $\pm$ 1.5\%, which is below the fraction in star-forming galaxies, and comparable to passive galaxies within the uncertainties. The fraction of more luminous AGN (lower panel of Figure \ref{fig:frac}) produces a similar pattern: the AGN fraction among PSBs is again below the star-forming sample in all bins, and more similar to passive galaxies, albeit with larger uncertainties.

Based on these individual detections, we conclude that we find no evidence for an excess of AGN activity in the recently quenched post-starburst population compared to other galaxies of comparable stellar mass and redshift. To probe lower levels of AGN activity, however, we must conduct a stacking analysis.

\begin{figure*}
 \includegraphics[width=2\columnwidth]{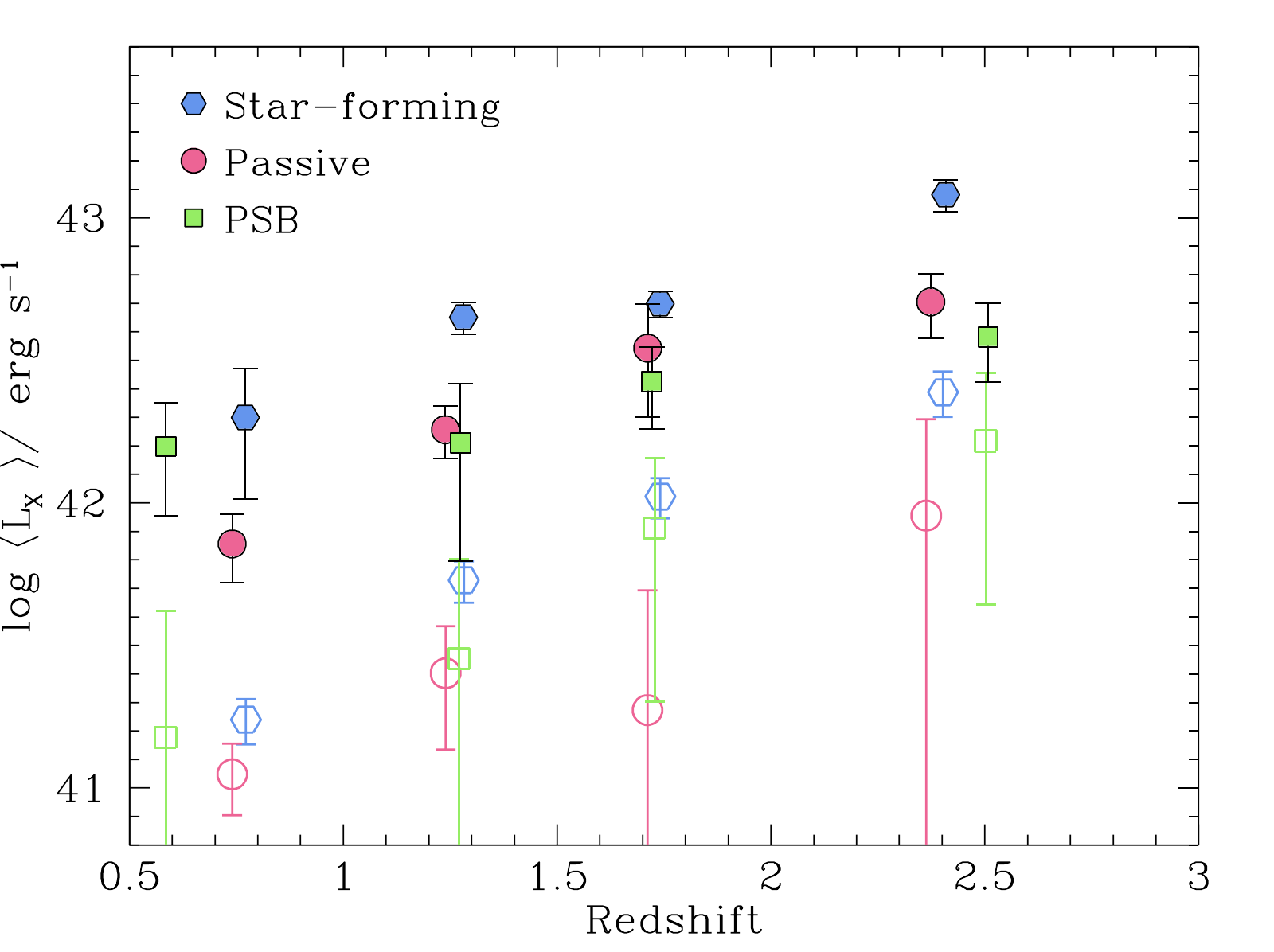}
 \caption{The mean $0.5-8$\,keV X-ray luminosity as a function of redshift for the three primary galaxy populations, determined for galaxies at high stellar mass ($M_\ast>10^{10.5}$ M$_{\odot}$). The unfilled points denote the average luminosity from a stacking analysis, excluding individual X-ray detections. The filled points combine stacked data 
 and individual detections to produce the overall mean X-ray luminosity for each population. 
Redshift bins of width $\Delta z=0.5$ are used in the range $0.5<z<2$, 
 with a single bin for the range $2<z<3$. Points are plotted at the mean redshift within each bin.
 }
 \label{fig:lumstack}
\end{figure*}

\section{X-ray Stacking analysis}
\label{section:stack}
The analysis of individual X-ray detections reveals no evidence for excess AGN activity in recently quenched galaxies, but 
would inevitably miss lower-luminosity activity. 
Such activity may be important if AGN are typically fading in the post-quenching phase \citep[e.g.,][]{Hopkins2012}, and lurking below the \textit{Chandra} detection limit.
To investigate lower-level AGN activity we therefore conduct a stacking analysis, taking advantage of the large number of galaxies within the \textit{Chandra} mosaic. The stacking is performed using the \textsc{CSTACK} tool version 4.5 
\citep{Miyaji2008},  to stack 
at the locations of
non-detected galaxies, after the removal of known X-ray sources.
The \textsc{CSTACK} algorithm is optimized to work with mosaiced \textit{Chandra} data, and performs a bootstrap analysis to obtain robust estimates of the mean count rate for a given population, and the associated uncertainties. Stacking is only performed at locations with 8 arcmin of the individual \textit{Chandra} field centres within the mosaic, matching the regions used for individual detections.
Count rates are determined in the $0.5-8$\,keV Chandra bands, and converted to flux using the \textit{Chandra} \textsc{PIMMS} tool, assuming a photon index $\Gamma=1.7$ and galactic absorption ($N_H=2.54\times10^{20} $cm$^{-2}$; 
\citealt{Dickey1990}).

\subsection{Average X-ray luminosity versus redshift}
\label{subsec:lum}

Figure \ref{fig:lumstack} presents the mean X-ray luminosity versus redshift for the three 
galaxy populations (star-forming, passive, and PSB). The analysis is restricted to galaxies at high stellar mass ($M_\ast>10^{10.5}$ M$_{\odot}$), to ensure a broadly complete galaxy sample as a function of redshift. The unfilled symbols represent the average luminosity from the stacking analysis, after excluding individual detections, with uncertainties from the \textsc{CSTACK} bootstrap procedure. The filled symbols represent the overall average X-ray luminosity, after combining the stacks with all individually detected X-ray sources. The small fraction of quasar-like AGN are removed, as outlined in Section \ref{section:quasars}.

Overall, we find that the mean X-ray luminosity is higher in the star-forming population at all redshifts. This is true for both the stacked and combined data.
Comparing star-forming and passive galaxies, the average offset in X-ray luminosity
is a factor $2.25\pm 0.24$  across the four redshift bins.
These results broadly confirm previous findings at similar redshifts, which also found higher AGN fractions and accretion rates in star-forming galaxies compared to passive galaxies \citep[e.g.,][]{Rodighiero2015, Aird2018, Carraro2020}. 
Our new finding, however, is that recently quenched galaxies (PSBs) appear to have average X-ray luminosities broadly similar to those of passive galaxies at the same redshift, but significantly below the average for star-forming galaxies, particularly at high redshift ($z>1$). Considering the mean offset between the four redshift bins, we find that PSBs show X-ray luminosities a factor $2.26\pm 0.37$ below the star-forming population on average.
If we restrict the comparison to the redshift range  $1<z<3$ (`cosmic noon'), removing the low redshift bin containing only 2 PSBs,  the average offset becomes a factor $2.6\pm 0.31$. 

We conclude, 
once again, 
that we find no evidence for any excess AGN activity associated with recently quenched galaxies at cosmic noon. On the contrary, the mean X-ray luminosity appears to be comparable to older passive galaxies, and significantly below that of star-forming galaxies of comparable redshift and stellar mass.

As a caveat, we note that the average X-ray luminosities may be underestimated for PSBs and passive galaxies if some of the host galaxies are misclassified.  A discussion of this issue is presented in Section \ref{section:sims}, where we conclude that the likely influence of this effect is small. Secondly, the X-ray luminosities will also be underestimated for heavily obscured AGN. To affect our conclusions, however,  we would require a higher fraction of heavily obscured AGN among the PSB population. A further discussion of this possibility is presented in Section \ref{section:missing}.

\subsection{Average X-ray luminosity versus star-formation rate}
\label{subsec:sfr}

\begin{figure}
  \includegraphics[width=\columnwidth]
 {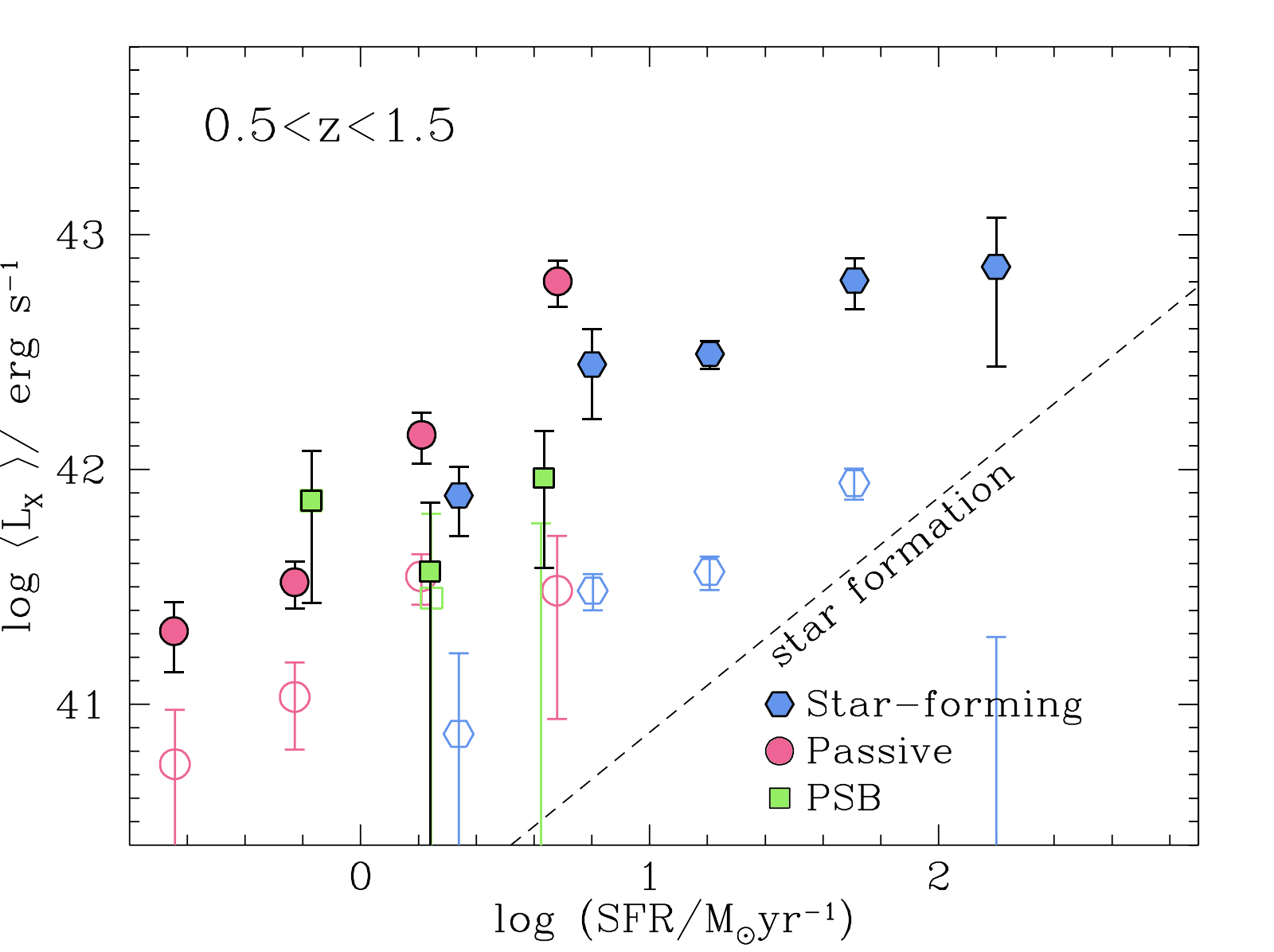}
  \includegraphics[width=\columnwidth]
 {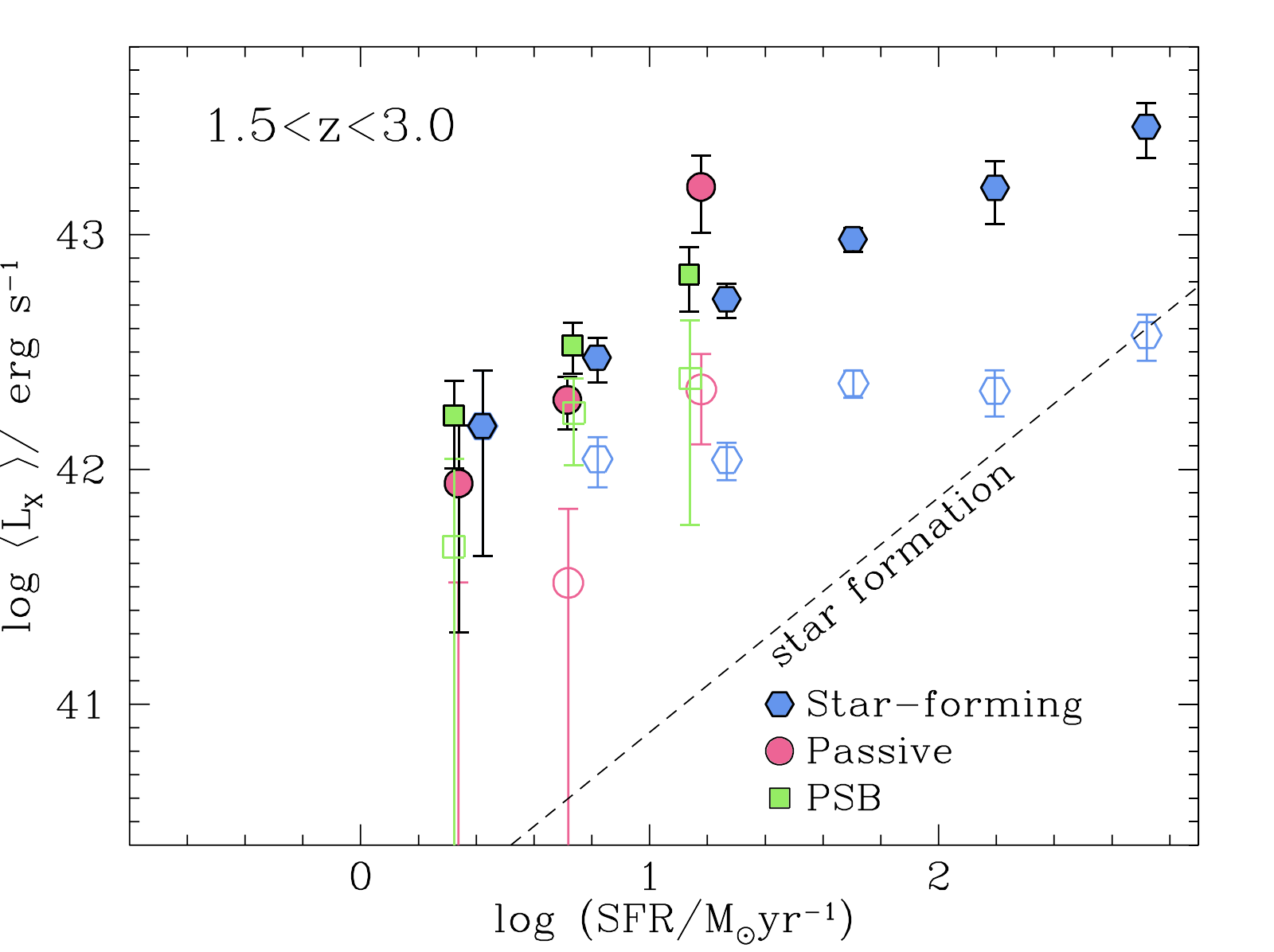}
 \caption{The mean X-ray luminosity as a function of star-formation rate for the three primary galaxy populations, determined for galaxies at high stellar mass ($M_\ast>10^{10.5}$ M$_{\odot}$).
  The unfilled points denote the average luminosity from a stacking analysis, excluding individual detections. The filled points combine stacked data 
 and individual detections.
  The dashed line denotes the expected relation if the X-rays are produced by processes related to star formation (e.g., massive X-ray binaries), from \protect\cite{Aird2017}. 
 }
 \label{fig:sfr}
\end{figure}

The results presented in Section \ref{subsec:lum} show a higher average X-ray luminosity in star-forming galaxies compared to passive systems, suggesting that AGN activity may be correlated with star formation. To explore this possibility further we conduct a stacking analysis for the three galaxy populations in bins of star-formation rate (SFR), with the individual SFR values determined from the Bayesian SED fitting procedure described in Section \ref{subsec:galaxies}.
Given the strong correlation of SFR with redshift, we conduct the stacking analysis in two redshift bins: $0.5<z<1.5$, and $1.5<z<3.0$. As before, we restrict the analysis to the most massive galaxies with $M_\ast>10^{10.5}$ M$_{\odot}$. The results are shown in Figure \ref{fig:sfr}, where as before we present the stacking results for undetected galaxies (unfilled) and the overall average X-ray luminosity (filled points) after including individual detections.

In both redshift bins we find a strong correlation between average X-ray luminosity and SFR. 
These trends confirm previous findings 
\citep[e.g.,][]{Silverman2009, Carraro2020, Aird2019, Mountrichas2023}, suggesting that AGN activity is (on average) strongly correlated with star formation, and hence with the availability of molecular gas. 
Figure \ref{fig:sfr} also illustrates that the average levels of X-ray activity in all populations are significantly above the expectations from star formation activity.
The dashed line illustrates the fitted linear relation between X-ray luminosity
and SFR due to non-AGN processes (e.g., massive X-ray binaries, etc), as obtained by \cite{Aird2017}. We conclude that the mean X-ray luminosities are dominated by AGN emission.

Overall, we find that PSBs show relatively low average X-ray luminosities, entirely consistent with their reduced levels of star formation. At low redshift there is perhaps marginal evidence for enhanced AGN activity in the passive population at a given SFR,
but a larger study will be required to confirm this tentative finding.
We note the overlap in SFRs between the star-forming, PSB, and passive populations. This is due in part to the relatively wide range of stellar masses, even among galaxies with $M_\ast>10^{10.5}$ M$_{\odot}$ (see Figure \ref{fig:zmass}), and also because
galaxy classifications are determined from only the first two PCA supercolours, while all three supercolours are used to estimate physical properties (see Section \ref{subsec:galaxies}).

One important caveat is that blue AGN light may be affecting the determination of the SFRs, particularly for the more luminous Type 1 (unobscured) AGN.
A small subset of the most quasar-like AGN have already been removed (see 
Section \ref{section:quasars}), but contamination from lower luminosity systems
may still be affecting the correlation shown in Figure \ref{fig:sfr}.
This issue is explored further in Appendix \ref{section:appendix}, where we find that stricter luminosity cuts do inevitably impact the slope of the $L_{\rm X}$ versus SFR relation, but a strong correlation remains. 
We also note that contamination from blue AGN light is unlikely to affect the stacking results based on the larger samples of {\it undetected} X-ray AGN (unfilled symbols in Figure \ref{fig:sfr}), which also show a strong correlation between the average X-ray luminosity and SFR.
Further work will be required to determine the precise relationship between these parameters, which is beyond the scope of this present study. Our primary interest is the PSB population, which show no evidence for a significant excess in AGN activity, but instead show AGN activity consistent with their lower levels of star formation.

\subsection{Splitting quenched galaxies by age}
\label{subsec:age}
\begin{figure}
 \includegraphics[width=\columnwidth]{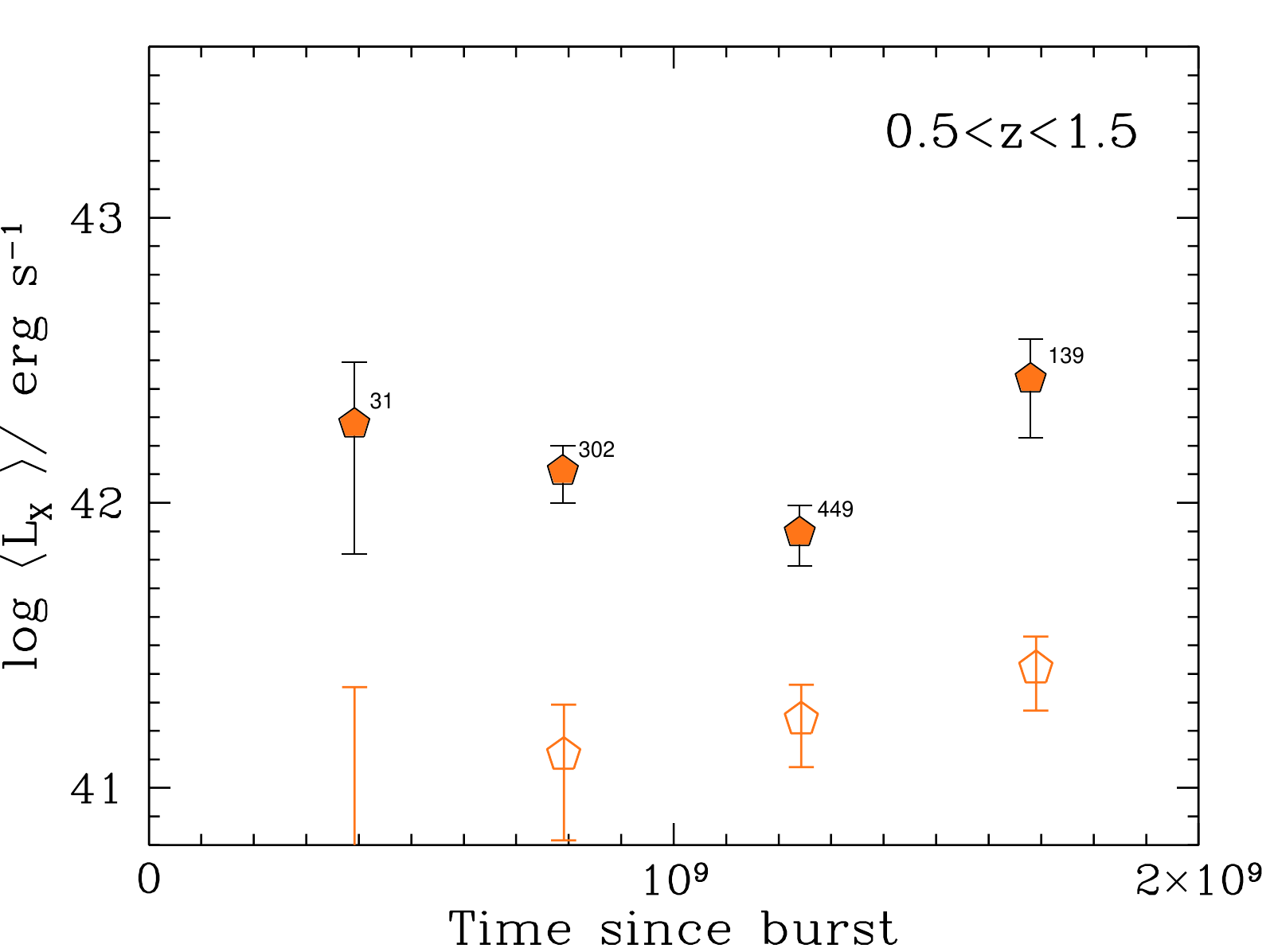}
  \includegraphics[width=\columnwidth]
 {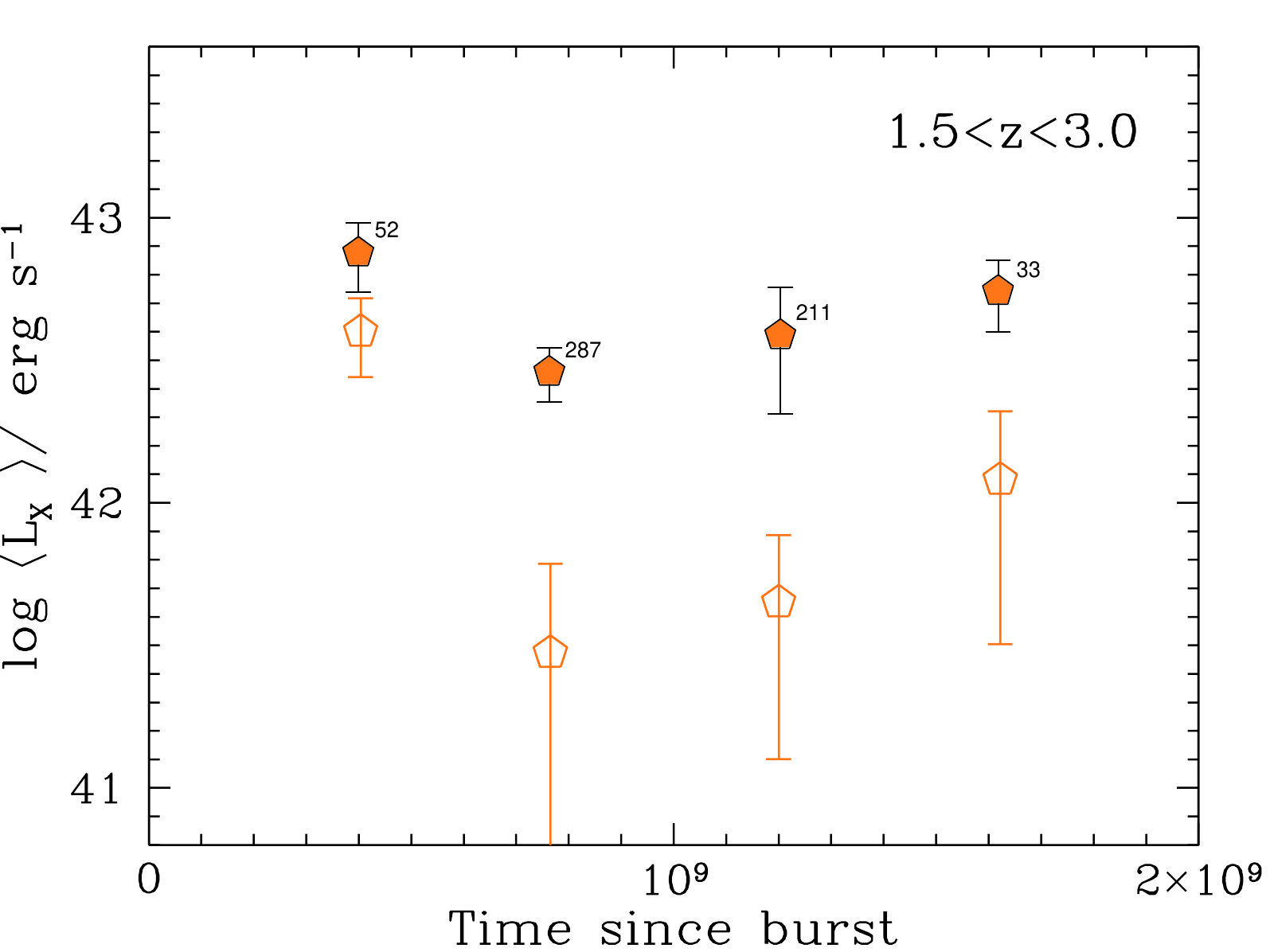}
 \caption{The mean X-ray luminosity for quenched galaxies (PSB and passive)
at high stellar mass ($M_\ast>10^{10.5}$ M$_{\odot}$), binned as a function of time since the last burst of star formation. Burst ages are determined from the SED fitting  outlined in Section  \ref{subsec:galaxies}. We caution that these ages are derived from photometric data, so may be subject to degeneracies between the age and strength of the burst (see Section \ref{subsec:age}). Four equally spaced age bins are used up to $2\times 10^9$ years, with the points plotted at the mean age within each bin.
  The unfilled points denote the average luminosity from a stacking analysis, excluding individual detections. The filled points combine stacked data 
 and individual detections to produce the overall mean luminosity for each population. The numbers denote the number of galaxies within each bin.}
  \label{fig:time}
\end{figure}

If AGN feedback is responsible for rapidly quenching massive galaxies we may expect the peak in AGN activity to occur a few hundred Myr after the peak in star formation, 
followed by a rapid decline as the black hole runs out of fuel \citep[e.g.,][]{Hopkins2012}. In the local Universe there is evidence that AGN activity peaks $\sim 250$ Myr after the peak in star formation \citep[][]{Wild2010, Davies2007, Yesuf2014}. 
During this declining phase, low-level AGN activity may be responsible for expelling any remaining gas in the galaxy, to hasten the transition to the red sequence \citep{Hopkins2012}.

To investigate AGN activity during the quenching phase, we combine the PSB and passive galaxy populations, and explore their mean X-ray luminosity as a function of the time since the most recent burst of star formation. The burst times were obtained 
from a Bayesian fit to the PCA supercolours for each galaxy, as described in Section \ref{subsec:galaxies}. Dividing the sample into two redshift bins, we then performed a \textit{Chandra} stacking analysis, following the procedure outlined in Section
 \ref{section:stack}. The results are presented in Figure \ref{fig:time}, where as before we display the mean X-ray luminosity from the stack of undetected galaxies (unfilled points), and after combining with individually detected X-ray AGN, to obtain the overall average luminosity (filled points). The results
 suggest no evidence for a significant trend in the average AGN luminosity with time since quenching. The modest upturn in the youngest bin at high redshift ($z>1.5$) can largely be explained by the enhanced average SFRs in this bin (see Figure \ref{fig:sfr}). 
 We note, however, that we have relatively few galaxies in the youngest time bins ($t_{\rm burst}<500$\, Myr),  within which the mean ages are $\sim 400$\, Myr. 
  Future refinement in the PCA selection criteria may also allow us to identify more galaxies at these earlier phases of quenching (Harrold et al., in preparation).
  The ages obtained from the photometric PCA analysis are also likely to be uncertain, due
 to degeneracies between the age and the strength of the burst \citep[see][]{Wild2020}.
We also note that our results appear to differ from the work of \cite{Ni2023}, 
who find evidence for a decline in AGN activity with age for quiescent galaxies at $z<1$, with age determined from the strength of the 
 4000\AA\, break.
 
We conclude that  a definitive assessment of high-z AGN activity as a function of time since quenching will require larger spectroscopic samples of PSBs and passive galaxies, to allow a more precise determination of individual SFHs.

\section{Quantifying the impact of AGN light on galaxy classification}
\label{section:sims}

In this section we consider the possibility that the presence of bright AGN may alter the photometric  classification of galaxies. In particular, the presence of a relatively blue AGN  could lead to a quenched galaxy being incorrectly classified as a star-forming system. We develop a simple model to investigate the impact of this effect,  adding AGN light to the SED of a typical PSB before reevaluating the PCA classification.
This process was repeated for a range of AGN accretion rates and nuclear 
reddening values. The key steps in the modelling are as follows:

\begin{figure}
 \includegraphics[width=\columnwidth]{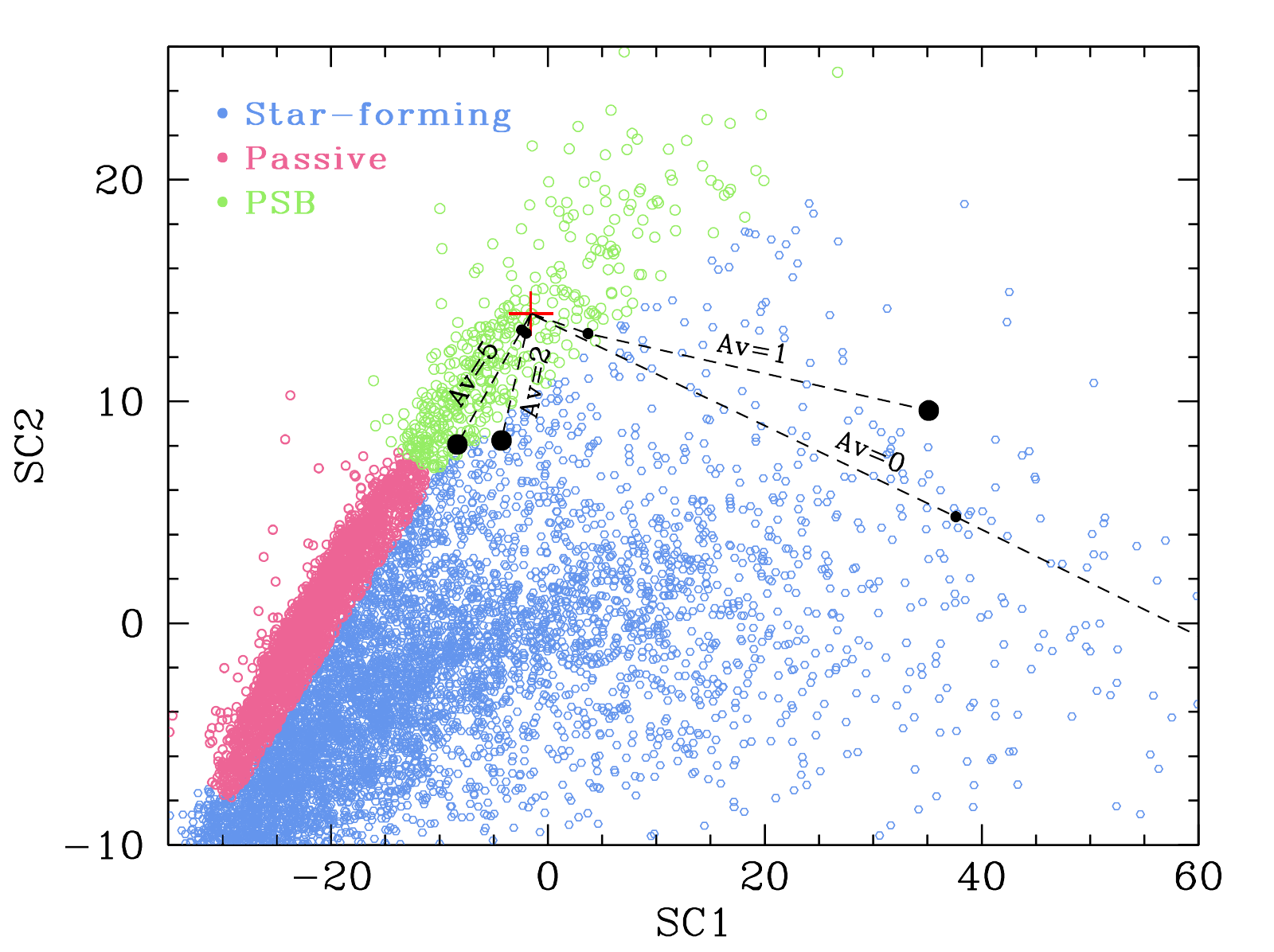}
  \includegraphics[width=\columnwidth]{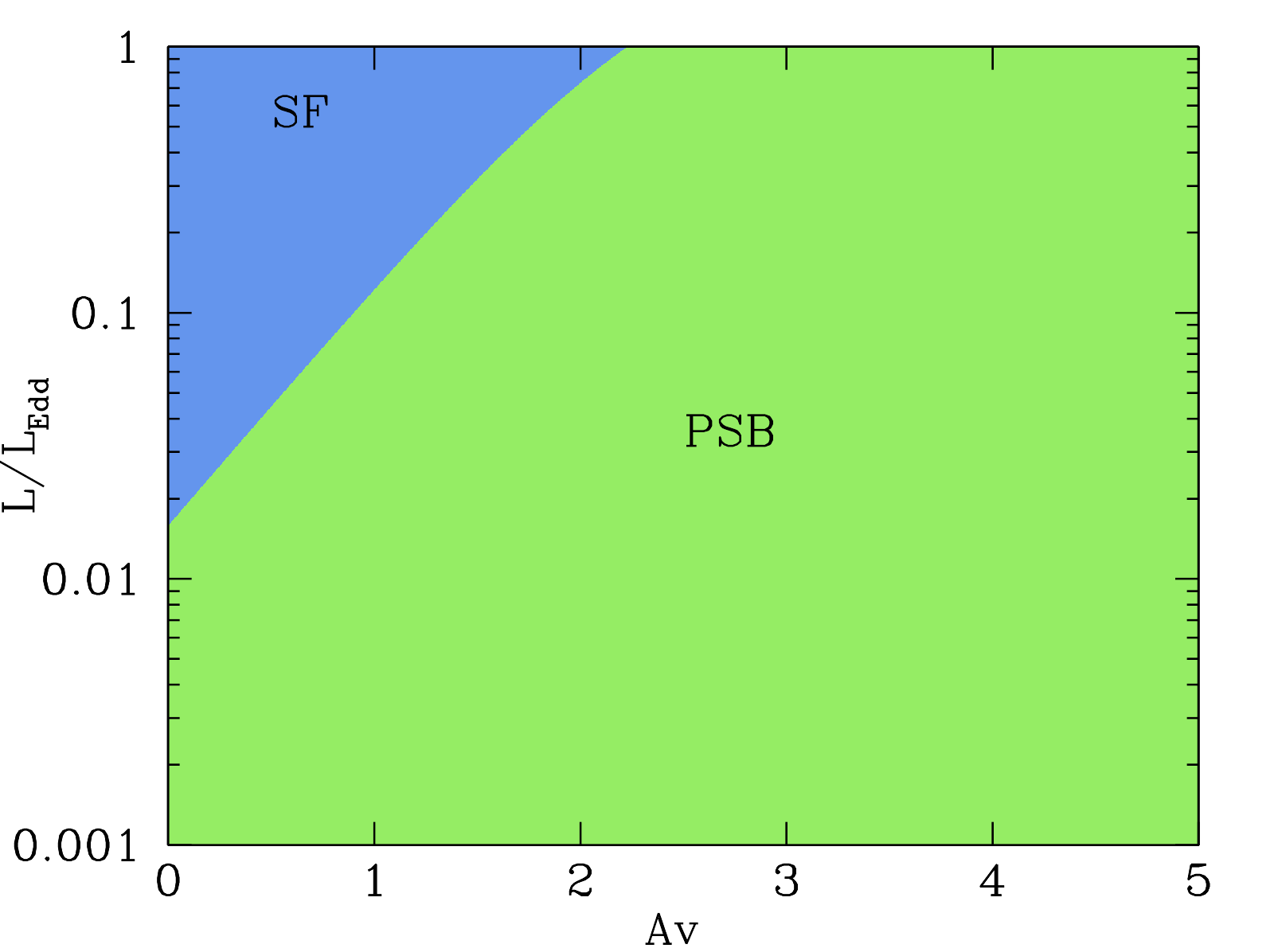}
 \caption{The upper figure shows the PCA `supercolour' diagram, used to classify galaxies into star-forming, passive and PSB categories \citep{Wild2014, Wilkinson2021}. The black curves show the impact of adding AGN light to a typical PSB (initially located at the red cross). Four curves are shown, representing AGN reddened by dust with  $A_{V}=0, 1, 2$\, and $5$. The small and large black points (linked by dashed lines) represent accretion at 
 $L/L_{\rm Edd}=0.1$\, and $1.0$ respectively. The lower figure shows the impact on galaxy classification for this typical object for a wider range of $A_{V}$ and $L/L_{\rm Edd}$.  }
  \label{fig:SC}
\end{figure}

\begin{figure}
 \includegraphics[width=\columnwidth]{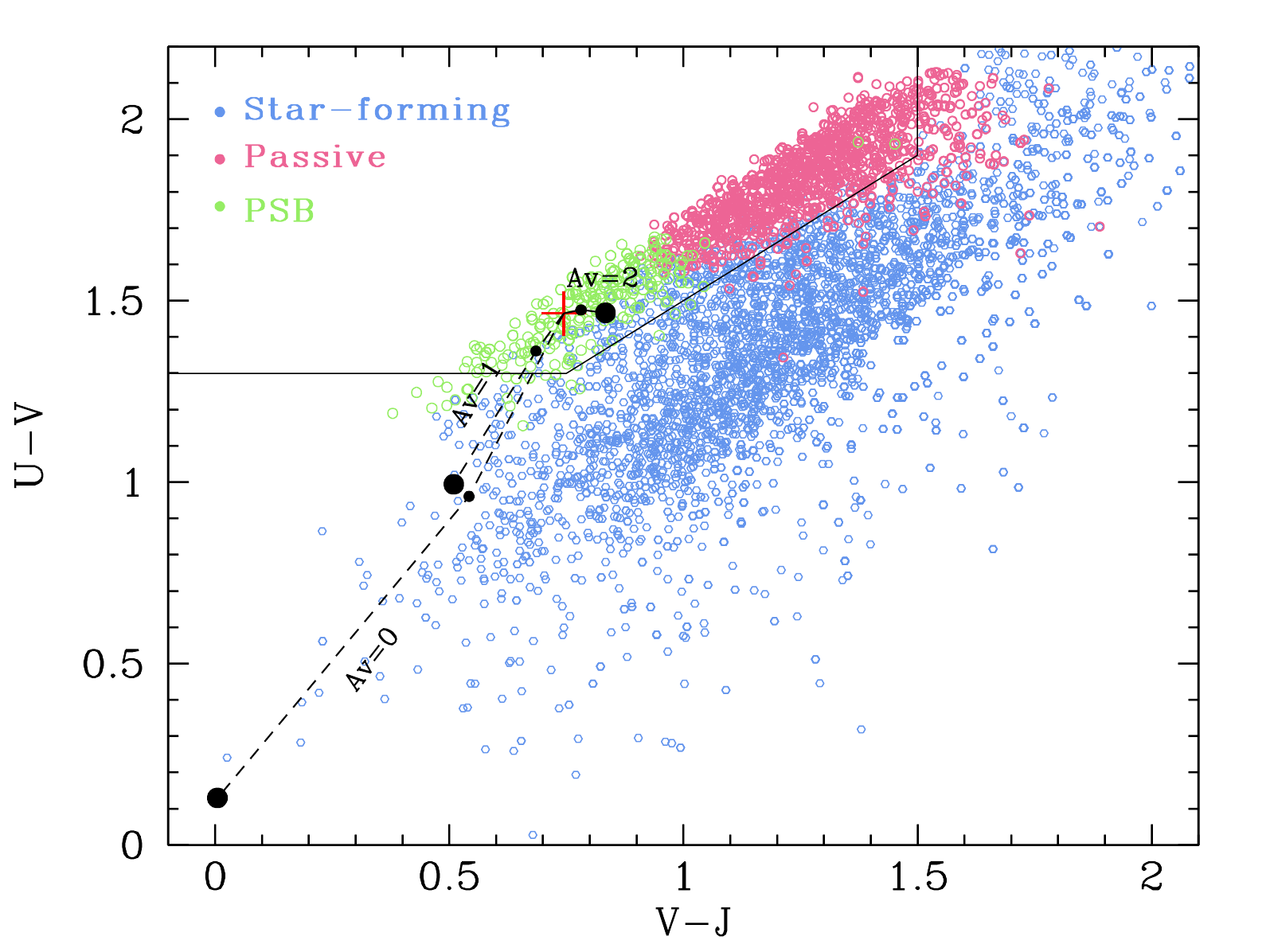}
 \caption{This figure shows the rest-frame UVJ colour-colour diagram, with the classification boundaries from \protect\cite{Whitaker2012}. Galaxies classified by the PCA supercolour technique are overlaid for comparison. Rest-frame UVJ colours are determined from the best-fitting supercolour SED for each galaxy.
The black curves show the impact of adding AGN light to a typical PSB (initially located at the red cross). Three curves are shown, representing AGN reddened by dust with  $A_{V}=0, 1$\, and $2$.  For clarity, curves with stronger dust reddening ($A_{V}>2$) are not shown, as these lie entirely within the PSB region and strongly overlap the curve at $A_{V}=2$.
The small and large black points (linked by dashed lines) represent accretion at 
 $L/L_{\rm Edd}=0.1$\, and $1.0$ respectively. 
 }
  \label{fig:UVJ}
\end{figure}

\medskip

\begin{itemize}
    \item To model the multiwavelength emission from the  AGN we use the broad-band  radio-quiet quasar SED from \cite{Elvis1994}, supplemented
    by more detailed UV to mid-IR SED from \cite{Assef2010}, matching the two SEDs
    at $10$ microns.
    \item We normalise the AGN SED 
    by assuming the bolometric luminosity
is produced by a black hole radiating at a fixed fraction of the Eddington luminosity. The black hole mass is estimated
assuming a black hole to stellar mass of $0.15$\% 
\citep[e.g.,][]{McLure2002, Haring2004}. A wide range of accretion rates are used, in logarithmic intervals from $L/L_{\rm Edd} = 10^{-3}$ to 
$L/L_{\rm Edd} = 1$. 

\item Most X-ray selected AGN are optically highly reddened \citep[e.g., ][]{Lawrence2010}, so we 
model the nuclear reddening with a range of values, from no reddening  
up to $A_V=5$.
The reddening laws for AGN are complex and poorly understood, 
 with studies suggesting a range of properties from steep SMC-like extinction to 
flatter `grey dust' \citep[see discussion within][]{Li2007}. For simplicity, we therefore adopt a Milky Way reddening law with $R_V=3.1$, which is broadly intermediate between those  extremes.  

\item We add the resulting AGN SEDs to the SED of a PSB, to investigate the impact on galaxy classification. We selected a typical PSB  at $z\simeq1.5$, with $M_\ast = 10^{11}$ M$_{\odot}$, with PCA supercolours colours close to the mean for the PSB population. We then added the quasar light, with a range of accretion rates ($L/L_{\rm Edd}$) and reddening ($A_V$), and feed the resulting modified SEDs through the PCA analysis (Section \ref{subsec:galaxies}),    to determine stellar masses, SFRs, and galaxy classifications. A range of alternative PSBs were tested, which yielded very similar conclusions.

\end{itemize}

The results from this simple modelling suggest that the stellar mass is relatively robust to low/moderate AGN activity. At accretion rates $L/L_{\rm Edd}<0.1$, reflecting the majority of AGN in our sample, the impact on the stellar mass is predicted to be $<20$\%. Stellar masses change by a factor of two or more only at very high Eddington rates ($L/L_{\rm Edd}>0.8$). 

The star-formation rates (SFRs)  of quenched galaxies are  much more sensitive to AGN light, though only at low levels of AGN obscuration. At moderate accretion rates ($L/L_{\rm Edd}>0.1$) and low obscuration ($A_V\ltsimeq1$) the SFRs for PSBs and passive galaxies can be boosted by a factor of 2 or more. For heavily reddened AGN ($A_V\gtsimeq 2$), however,  the impact on the SFRs is negligible.

The impact on the location of a typical PSB in the supercolour and UVJ and diagrams is illustrated in Figures \ref{fig:SC} and \ref{fig:UVJ}, where the colours are determined from the best-fitting PCA supercolour templates described in
Section \ref{subsec:galaxies}. For Type 1 AGN at low levels of obscuration ($A_V\ltsimeq1$), we find that PSBs could be misclassified as star-forming galaxies, even at relatively low accretion rates. For heavily reddened AGN
($A_V>2$) the 
classifications appear to be robust to the presence of AGN.

To address the potential impact of the AGN contribution we adopt two approaches, reflecting two broad regimes:

\begin{enumerate}
    \item Our simulations suggest that galaxy properties and classifications are likely to be most unreliable at relatively high accretion rates ($L/L_{\rm Edd}>0.1$) and at low levels of obscuration ($A_V\ltsimeq1$). For a typical passive galaxy or PSB in our sample at $M_\ast\simeq 10^{11}$ M$_{\odot}$, this accretion rate corresponds to an 
 X-ray luminosity 
$L_{\rm X} \simeq 10^{44}$ erg s$^{-1}$ ($0.5-8$\,keV), assuming a bolometric correction factor $k_{\rm bol}=18$ to scale from the X-ray to total AGN luminosity (based on the SED described above). As described in Section \ref{section:quasars}, we have already removed 
all X-ray AGN above this X-ray luminosity that also display evidence for point-like profiles in our 
 $K$-band imaging (\texttt{CLASS\_STAR}\,$>0.95$). This cut removed ($\sim 8 $\%) of X-ray sources, which we assume would have the least reliable classifications, stellar masses, and SFRs.  
 As discussed in Appendix \ref{section:appendix}, this cut has only a very minor impact on the key results presented in this work.  Stricter or more relaxed exclusion criteria also do not appear to have a major impact on our conclusions.
 
\item To investigate the potential misclassification of other quenched galaxies that do not meet the quasar-exclusion threshold, we can use Figure \ref{fig:SC} to estimate the fraction of PSBs and passive galaxies that are likely to be misclassified. Previous work has suggested that approximately 30\% of AGN  are unobscured (`Type 1') systems with $A_V\ltsimeq1$ \citep{Lawrence2010}. According to Figure \ref{fig:SC}, such AGN could be missing from our quenched populations at accretion rates $L/L_{\rm Edd}\gtsimeq0.02$.
If we assume this missing $30$\% have similar
X-ray luminosities to the detected population, 
we can perform a basic correction, by artificially increasing the contribution from detected AGN in quenched galaxies with
$L/L_{\rm Edd}\gtsimeq0.02$ by $50$\%, and subtracting the equivalent luminosity from the star-forming populations.
We can then calculate the impact on the mean X-ray luminosities, e.g., as presented in Figure \ref{fig:lumstack}.
 We note that there should be negligible impact on the stacks of \emph{undetected} objects (unfilled points in Figure \ref{fig:lumstack}), which would typically have Eddington rates well below 1\%.
Performing this simple correction boosts the average luminosity for the PSBs and passive galaxies in Figure \ref{fig:lumstack} by $0.09$ and $0.14$ dex respectively, and reduces the average luminosity for star-forming galaxies by approximately $0.02$ dex. 
The passive and PSB populations therefore remain consistent with each other, and both remain significantly below the average X-ray luminosities of  star-forming galaxies.

\end{enumerate}

We conclude that a potentially significant fraction of passive galaxies and PSBs hosting luminous AGN may be missing from photometric samples, if those AGN are relatively unobscured ($A_V\ltsimeq1$).
Based on our simple analysis, however, we find this missing population is unlikely to affect our primary conclusion that we find no evidence for an excess of AGN activity in recently quenched galaxies.

\section{Caveat on heavily obscured AGN}
\label{section:missing}

Our analysis suggests no excess of X-ray AGN activity in massive, photometrically selected PSBs at cosmic noon, both from individual detections (Section \ref{section:detections}) and a stacking
analysis as a function of redshift (Section \ref{subsec:lum}). Instead, the PSBs appear to have low levels of AGN activity entirely consistent with their low SFRs (Section \ref{subsec:sfr}). However, it is important to recognise that heavily obscured AGN may be missing from our analysis. 

\begin{figure}
 \includegraphics[width=\columnwidth]{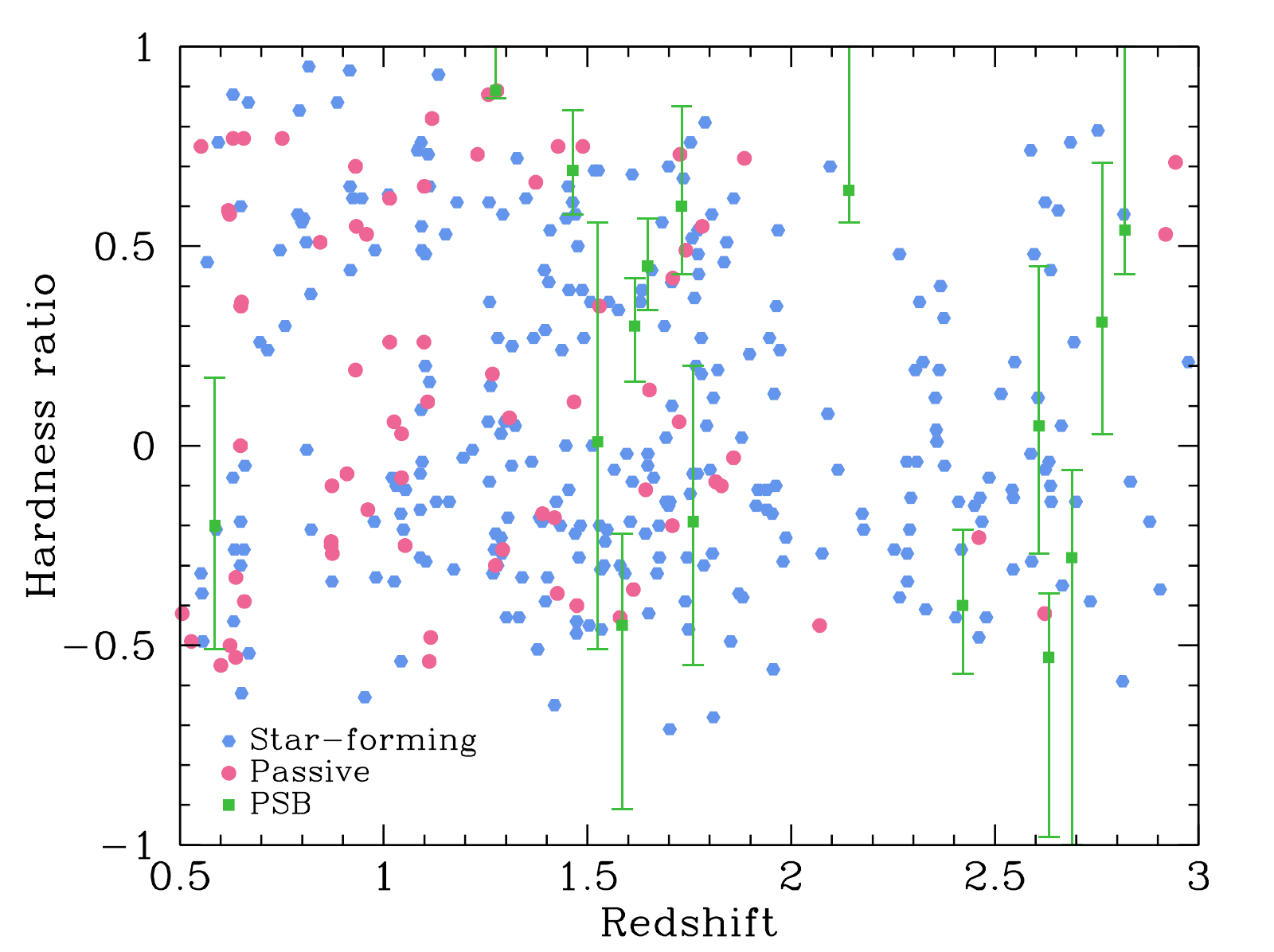}
 \caption{
The \textit{Chandra} X-ray hardness ratios as a function of redshift for the high-mass ($M_\ast>10^{10.5}$ M$_{\odot}$) 
star-forming, passive, and PSB galaxies detected as X-ray sources. For clarity the $1\sigma$ uncertainties are displayed for the PSBs only.
Hardness ratios and their uncertainties were obtained from the catalogue of \citet[][]{Kocevski2018}. We find no evidence that the PSBs show a significantly different distribution of hardness ratios compared to passive or star-forming galaxies, according to a K-S test.
}
  \label{fig:hr}
\end{figure}

Photoelectric absorption preferentially removes low-energy photons, so the bias against absorbed AGN diminishes somewhat towards high redshift. Nevertheless, X-ray studies may miss the most heavily obscured AGN, particularly if the obscuring gas is Compton-thick  (i.e.,  $N_H>\sigma_T^{-1} \simeq 1.5\times 10^{24}\,$cm$^{-2}$). 
Up to 50\% of Type 2 AGN may be Compton-thick \citep[][]{Risaliti1999, Lawrence2010}, and we expect supermassive black holes to be heavily obscured during their primary growth phase \citep[e.g.,][]{Fabian1999, Hopkins2006}. 
Our stacking analysis (e.g., Figure \ref{fig:lumstack}) would therefore inevitably underestimate the intrinsic AGN power within the galaxy populations. To affect our primary conclusions, however, would require a higher fraction of obscured AGN among the PSB population.
Investigating this possibility would require a more detailed analysis of the X-ray spectra as a function of redshift and galaxy type, which is beyond the scope of this work, and would likely require larger samples.  However, we note that the distribution of X-ray hardness ratios does not indicate significant differences between the three populations. In Figure \ref{fig:hr} we display the hardness ratios as a function of redshift for the star-forming, passive, and PSB galaxies. The hardness ratios  were obtained from the catalogue of \citet[][]{Kocevski2018}, who used a Bayesian analysis to allow for the Poisson nature of the source counts, 
with the hardness ratio defined as $HR = (H-S)/(H+S)$, where $H$ and $S$ reflect the counts in the $2-10\,$keV and $0.5-2\,$keV bands respectively. 
A 2-sample Kolmogorov-Smirnov (KS) test is unable to reject the null hypothesis that the hardness ratios of either the passive and PSB populations are drawn from the same underlying distribution as the star-forming sample ($p$-values > 0.05).
However, the sample of X-ray detected PSBs is small and the uncertainties on the hardness ratios are significant. Larger samples may therefore be required to determine whether highly obscured AGNs are more common in this population.
A complementary approach would be to search for the hot dust emission expected from heavily enshrouded AGN, e.g., using the Mid-Infrared Instrument \citep[MIRI;][]{Rieke2015} on board the James Webb Space Telescope (\textit{JWST}).

\section{Discussion}
\label{section:discussion}

\subsection{Comparison with previous studies of AGN in recently quenched galaxies}

Our results provide strong evidence that recently quenched massive galaxies at cosmic noon show no apparent excess of X-ray AGN activity compared to galaxies at other stages in their evolution. These findings are in good agreement with studies in the more local Universe. For example, \cite{Lanz2022} study a sample of 12 shocked PSBs at redshifts $z<0.2$, finding evidence for weak AGN activity in approximately half the sample. Most of their AGN have low X-ray luminosities
($L_{\rm X} < 10^{42}$ erg s$^{-1}$, $2-10$\,keV), consistent with our stacked results at $z<1$ (Figure {\ref{fig:lumstack}). \cite{Lanz2022}} conclude that such low-level activity would be insufficient to quench these galaxies, and hence that AGN are "along for the ride", and simply tracing the availability of gas. \cite{French2023} find evidence for fading AGN in 6/93 local PSBs, based on extended light echos observed in 
[OIII]\,$\lambdaup$5007. They conclude that the AGN in these galaxies are generally in a very low luminosity phase, but are visible in a luminous phase approximately 5\% of the time. These results are again in good agreement with our findings at higher redshift.

The launch of \textit{JWST} has recently allowed detailed spectroscopic studies of higher-redshift quenched/quenching galaxies at $z\simeq2-4$. So far a number of studies have reported evidence for AGN activity in many of these galaxies, typically based on line ratio
diagnostics
\citep[e.g.,][]{Belli2024, Davies2024, D'Eugenio2024, Wu2024}. 
The samples are relatively small so far, however, so it is currently unclear 
if the implied prevalence of AGN is in tension with our \textit{Chandra} X-ray study.
Shocks can also give rise to AGN-like line ratios, and shocks are found to be common in low-redshift PSBs 
\citep[e.g.,][]{Alatalo2015}.
Larger samples and additional AGN diagnostics may be needed to determine if the AGN activity seen by 
\textit{JWST} 
is significantly enhanced compared to other galaxies with similar stellar mass and redshift.
Given the rapid nature of AGN variability \citep[e.g.,][]{Hickox2014}, 
it is also possible that line ratio diagnostics overestimate the fraction of AGN undergoing nuclear activity compared to X-ray studies, as line emission can arise from light echos caused by previous AGN episodes \citep[e.g., see][]{French2023, Sell2014}, while X-ray emission is effectively an instantaneous probe of black hole activity.

\subsection{Reconciling the low AGN prevalence with high-velocity outflows in PSBs}

In addition to direct evidence of AGN activity, there is strong evidence that many massive PSBs drive high-velocity interstellar outflows of $\sim$1000\,km\,s$^{-1}$ or higher \citep[e.g.,][]{Tremonti2007, Maltby2019, Man2021}. 
Such outflows are difficult to explain without invoking AGN activity, particularly in cases where outflows are detected long after the star formation has been quenched \citep{Taylor2024}. We may therefore ask how our low AGN fraction can be reconciled with the presence of these powerful outflows.

Our study suggests that approximately $5$\% of massive PSBs are detected as luminous AGN at any given time,
 with average
X-ray luminosities of approximately L${\rm _X}\sim 5\times10^{43}$ erg s$^{-1}$ (0.5-8 keV), corresponding to accretion rates of  $L/L_{\rm Edd}\sim 0.05$.
The remaining PSBs have average X-ray luminosities more than an order of magnitude below this level (Figure \ref{fig:lumstack}). These findings can be reconciled with the strong prevalence of outflows if the luminous phase of AGN activity is short-lived.
During the luminous "on" phase, the implied bolometric AGN power 
would be at least an order of magnitude higher than the 0.5-8 keV X-ray luminosity, i.e. $P_{\rm AGN} \sim 5\times10^{44}$ erg s$^{-1}$. 
We can compare this AGN power with the observed kinetic power in the outflowing gas. For an outflow with a  velocity of $\sim$1000\,km s$^{-1}$, assuming a total mass outflow rate of $\dot{M}\sim 20$\,M$_{\odot}$\,yr$^{-1}$ \citep[e.g.,][]{Wu2024}, we obtain an outflowing kinetic power of approximately $P_{\rm Outflow} = \frac{1}{2} \dot{M} v^2 \sim 6\times10^{42}$ erg s$^{-1}$. 
Thus only $1-2$\% of the bolometric power generated by the AGN would be sufficient to drive the outflow, if the AGN is strongly coupled to the surrounding gas.
Theoretical predictions suggest a coupling efficiency of $\sim$5\% for AGN driving outflows when close to the Eddington limit, with less efficient coupling at lower accretion rates \citep[e.g., see][]{King2010, Cicone2014}. 
Thus the occasional episodic AGN activity implied by our X-ray detection rate could plausibly launch outflows similar to those observed. If the AGN remains "on" for $\sim 1$\,Myr (say), at a velocity of 
$\sim$1000\,km s$^{-1}$ the outflow would reach a radius of $\sim 1$\,kpc, comparable with the stellar half-light radii \citep{Almaini2017}. If the AGN then switches off, the relic outflow could then be visible for up to $\sim10$\,Myr. Thus, with a relatively short AGN duty cycle, the outflow could be visible for significantly longer than the luminous X-ray AGN. This scenario could also explain the apparently high fraction of high-z PSBs showing evidence for high-ionisation signatures \citep[e.g.,][]{Bugiani2025}, despite low levels of X-ray activity.

\begin{figure}
 \includegraphics[width=\columnwidth]{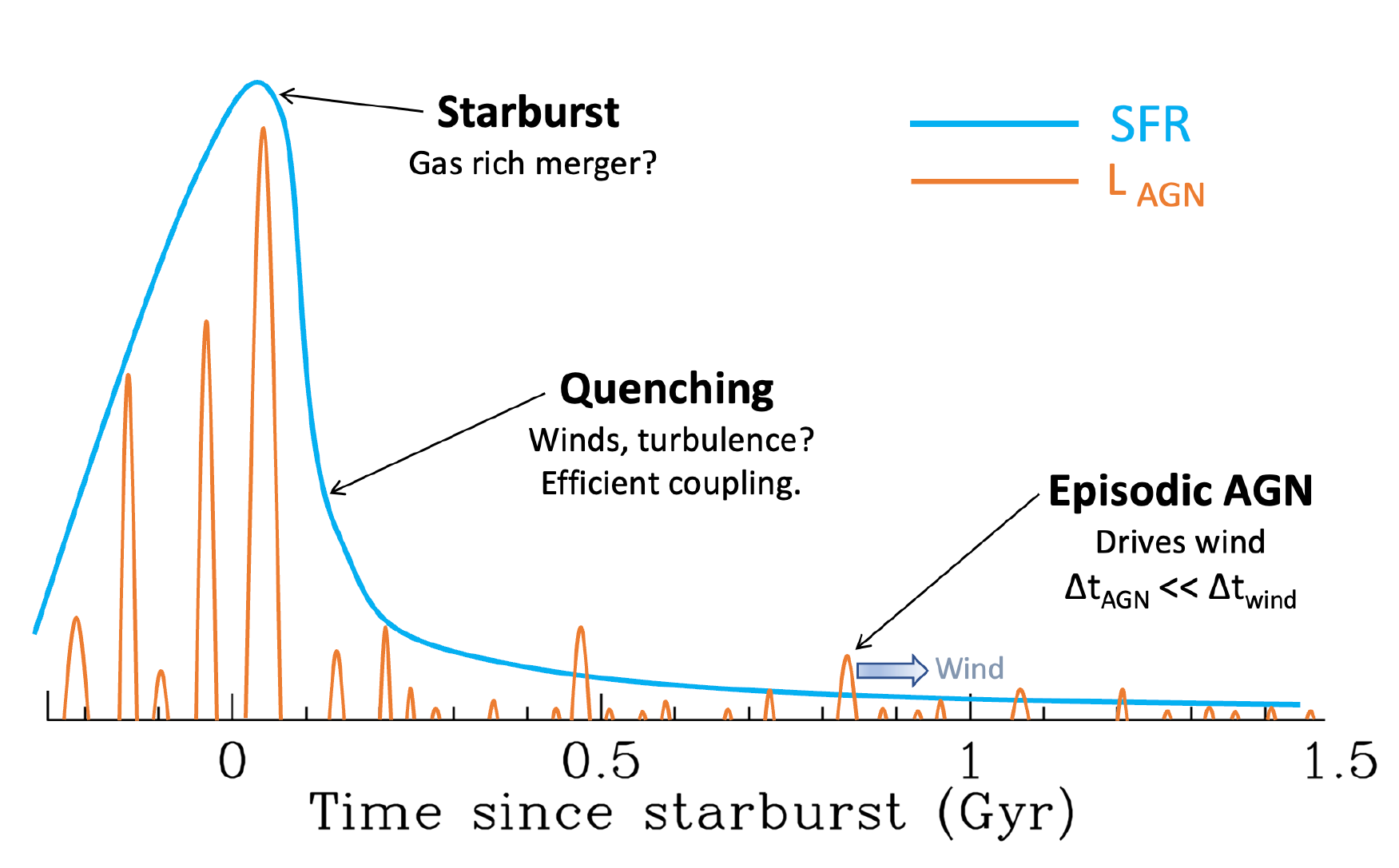}
 \caption{An illustration of a 
 plausible scenario for the co-evolution of AGN activity and star formation in a recently quenched galaxy, consistent with our \textit{Chandra} observations.
 AGN activity broadly traces the SFR (displayed with arbitrary scaling), but
there is no excess of X-ray activity associated with quenching. 
Occasional stochastic bursts of nuclear activity (visible $\sim 5$\% of the time) are sufficient to couple strongly to the surrounding gas and drive a powerful AGN-driven wind. If the AGN duty cycle is short, the relic outflow can persist for significantly longer than the visible X-ray AGN. Continued bursts of activity may eventually deplete the gas reservoir.
 }
  \label{fig:scenario}
\end{figure}

We therefore propose a plausible scenario illustrated in Figure \ref{fig:scenario}, in which AGN activity broadly traces the SFR, consistent with our findings, but in bursts of short-lived activity. During a major episode of star formation the bursts of AGN activity are likely to be more luminous, and may contribute to the rapid quenching of star formation. The primary quenching event may have been caused by the cumulative energy input from AGN \citep[e.g.,][]{Piotrowska2022, Harrison2024}, or the more explosive feedback predicted from a combination of AGN winds and Compton heating \citep[e.g.,][]{Choi2015, Hopkins2016}. 
High-velocity winds may also be driven by extreme densities of star formation, without input from AGN activity \citep[e.g.,][]{Diamond-Stanic2012, Davis2023}.
However, there is evidence that AGN activity can significantly boost the velocity of interstellar outflows during the starburst phase \citep[e.g.,][]{Talia2017}.
Whatever the cause of the initial quenching event, our results suggest no evidence for a strong excess of AGN activity continuing into the post-starburst phase,  in apparent contrast to the expectations from some
theoretical models \citep[e.g.,][]{Hopkins2012}. A reservoir of cold gas may then persist within the central regions of the galaxy, as observed in many PSBs \citep[e.g.,][]{Rowlands2015, French2015, Suess2017,Umehata2025}, 
sufficient to fuel  low-level star-formation and occasional stochastic bursts of luminous AGN activity. These occasional bursts are short-lived, but can drive an outflow that is visible for significantly longer than the luminous AGN phase. These AGN may also inject turbulence, to help suppress star formation
\citep[e.g.,][]{Alatalo2015, Smercina2022, Girdhar2024}, while the outflows eventually deplete the central gas reservoir.   The galaxy itself would therefore remain quiescent, unless the system is rejuvenated, e.g., by another gas-rich merger. The galaxy gradually accretes satellites through minor mergers, until the host dark matter halo is sufficiently massive to host a static hot X-ray reservoir. At this point, radio-mode feedback may act to suppress a cooling flow and maintain quiescence in the central galaxy \citep[e.g.,][]{Croton2006}. This eventual outcome would be consistent with the observed large-scale clustering of massive PSBs \citep{Wilkinson2021}, which appear to reside in massive dark matter halos 
$M_{\rm halo}\simeq 10^{13}$\,M$_{\odot}$, and are likely to become
the central galaxies within galaxy clusters by the present day.

\subsection{Discussion on PSB visibility timescales}

As a final caveat, we note that we cannot rule out the possibility
that AGN activity is enhanced in the earliest stages of quenching.
 Our PSB samples  are photometrically selected, and can be identified up to $\sim$1\,Gyr after the quenching of a strong burst  \citep[see][]{Wild2020}. It is therefore possible that enhanced AGN activity is present in the earliest stages 
 \citep[e.g., up to 250\,Myr after the burst;][]{Wild2010, Hopkins2012}, and potentially washed out given the effective time resolution of our selection technique. We attempted to separate the PSBs by age in  Section \ref{subsec:age}, finding no obvious trends, but noted the strong degeneracies between age and burst fraction when using photometrically determined ages.

Future large spectroscopic samples observed  with \textit{JWST} may  provide the next breakthrough in this field, with the ability to constrain 
PSB star-formation histories to a precision of $\sim$100\, Myr  or better given sufficient spectral resolution and signal-to-noise 
\citep[e.g., see][]{Carnall2024}. With large spectroscopic samples, such studies may provide the time resolution required to study AGN activity (and outflows, shocks, and turbulence) as a function of time since the quenching event.

\section{Conclusions}

We present a \textit{Chandra} X-ray analysis of AGN activity in a large sample of photometrically classified massive 
    ($M_\ast>10^{10.5}$ M$_{\odot}$) galaxies
    in the UDS field, including over $4000$ galaxies 
   in the redshift range  $1<z<3$ (`cosmic noon'). The primary aim is to investigate AGN activity within the recently quenched class of post-starburst galaxies (PSBs), which may shed light on the early stages of galaxy quenching.

We find that     X-ray emitting AGN are
detected in 6.2 $\pm$ 1.5\% of massive PSBs in the redshift range $1<z<3$, 
with typical X-ray luminosities of 
${L\rm _X}\sim 5\times10^{43}$ erg s$^{-1}$ (0.5-8 keV), 
implying accretion at 
 $L/L_{\rm Edd}\sim 0.05$.
The detection rate lies between those of  star-forming and passive  galaxies 
(8.2 $\pm$ 0.5\% and 5.7 $\pm$ 0.8\%, respectively).
For the remaining undetected high-mass PSBs, a stacking analysis 
reveals weak AGN activity  
with average luminosities  of 
$\langle L_{\rm X} \rangle \sim 10^{41}-10^{42}$ erg s$^{-1}$,  implying very low 
average accretion rates with  $\langle L/L_{\rm Edd}\rangle \ltsimeq 10^{-3}$.
 For the population as a whole, combining stacks and individual detections, we find that PSBs
show average X-ray luminosities a factor $2.6\pm 0.3$ below those of star-forming galaxies 
of comparable stellar mass in the redshift range $1<z<3$,  but similar average X-ray luminosities to older passive galaxies.

We find evidence for a strong correlation between average X-ray luminosity and star-formation rate for all galaxy populations. We caution that this trend may be affected by optical/UV contamination from AGN light, but the results suggest that AGN activity, on average,  is driven by the availability of fuel. The low levels of AGN activity observed in PSBs therefore appear to be consistent with their relatively low levels of  star formation.

An important caveat is to recognise that photometric selection may miss quenched galaxies if they host relatively blue, luminous AGN.  We use a simple model to quantify the impact of this effect, showing that quenched galaxies may be misclassified as star-forming galaxies if they contain relatively unobscured AGN with low levels of nuclear reddening ($A_V\ltsimeq1$), particularly at high accretion rates
($L/L_{\rm Edd}>0.1$). 
A correction for this bias suggests that our key conclusions are not significantly affected. However, further work will be required to quantify the true fraction of AGN host galaxies potentially misclassified by photometric selection techniques.

Our X-ray analysis will underestimate the average AGN power if the galaxies contain a large population of heavily obscured (particularly Compton-thick) AGN. To affect our conclusions, however, would require a higher fraction of X-ray absorbed AGN among the PSB population. Hardness ratios suggest no significant excess of hard X-ray sources among the PSBs, but a future X-ray spectral stacking analysis with larger samples will help to fully quantify the impact of the highly absorbed population.

The low AGN fraction that we observe can be reconciled with the detection of high-velocity outflows in high-z PSBs, if the AGN duty cycle is short, i.e., the wind is visible for considerably longer than the luminous AGN phase. Energetically, the AGN that we observe in $\sim 5$\% of PSBs (accreting at $L/L_{\rm Edd}\sim 0.05$) are more than sufficient to drive the observed outflows.

Overall, we conclude that we find no evidence for excess X-ray AGN activity in recently quenched galaxies at cosmic noon. Instead, 
we find low levels of AGN activity, on average, consistent with the low rates of star formation in this transition population.
Episodic stochastic bursts of black hole activity may nevertheless help to maintain quiescence in this phase, e.g., by driving outflows, and clearing out gas from the nuclear regions.

\section*{Acknowledgements}
OA would like to thank Takamitsu Miyaji for helpful advice, and for providing the powerful CSTACK tool. OA acknowledges the support from STFC grant ST/X006581/1. VW acknowledges the support from STFC grant ST/Y00275X/1. We gratefully acknowledge support from the NASA Astrophysics Data Analysis Program (ADAP) under grant 80NSSC23K0495.
We would also like to thank an anonymous referee for their thoughtful and constructive comments.

\section*{Data Availability}

The data forming the basis of this work is available from public archives, further details of which can be obtained from the UDS web page (\url{https://www.nottingham.ac.uk/astronomy/UDS/}). A public release of the processed data, galaxy properties, and photometric redshifts is in preparation. Details can be obtained from the corresponding author.



\bibliographystyle{mnras}
\bibliography{almaini24} 




\appendix

\section{Robustness tests}
\label{section:appendix}

In this Appendix we present alternative versions of Figure \ref{fig:lumstack} and  Figure \ref{fig:sfr}, to investigate the impact of the cuts applied to remove quasar-like objects. By default, we removed objects that were both highly luminous ($L_{\rm X} > 10^{44}$ erg s$^{-1}$)
and point-like in our $K$-band imaging  (\texttt{CLASS\_STAR}\,$>0.95$), for which the galaxy classifications, stellar masses, and SFRs may be unreliable, as explained in Section
\ref{section:quasars}
and Section \ref{section:sims}.
These cuts removed 32 out of 407 X-ray sources from our massive galaxy sample.

\begin{figure}
 \includegraphics[width=\columnwidth]{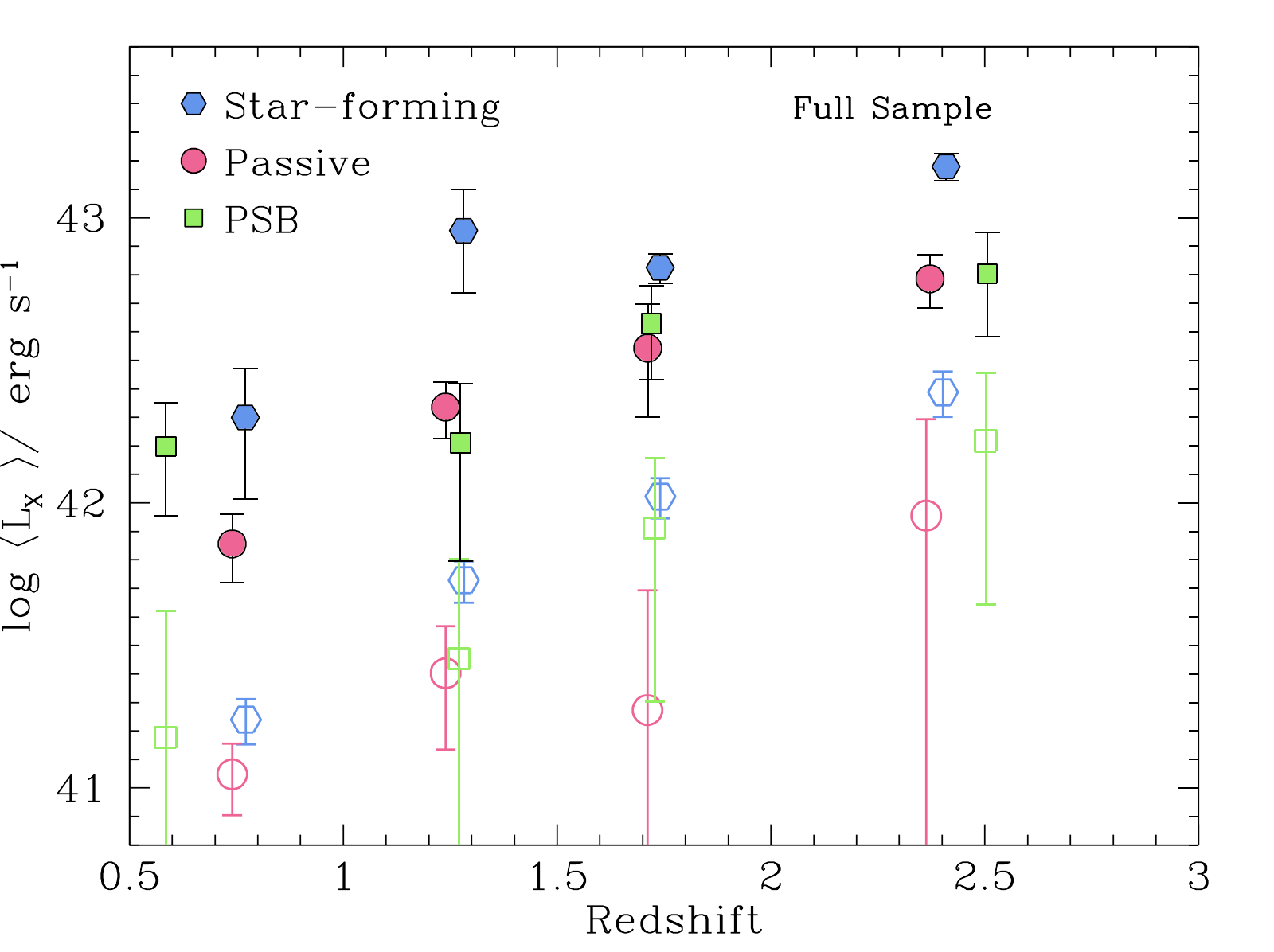}
 \includegraphics[width=\columnwidth]{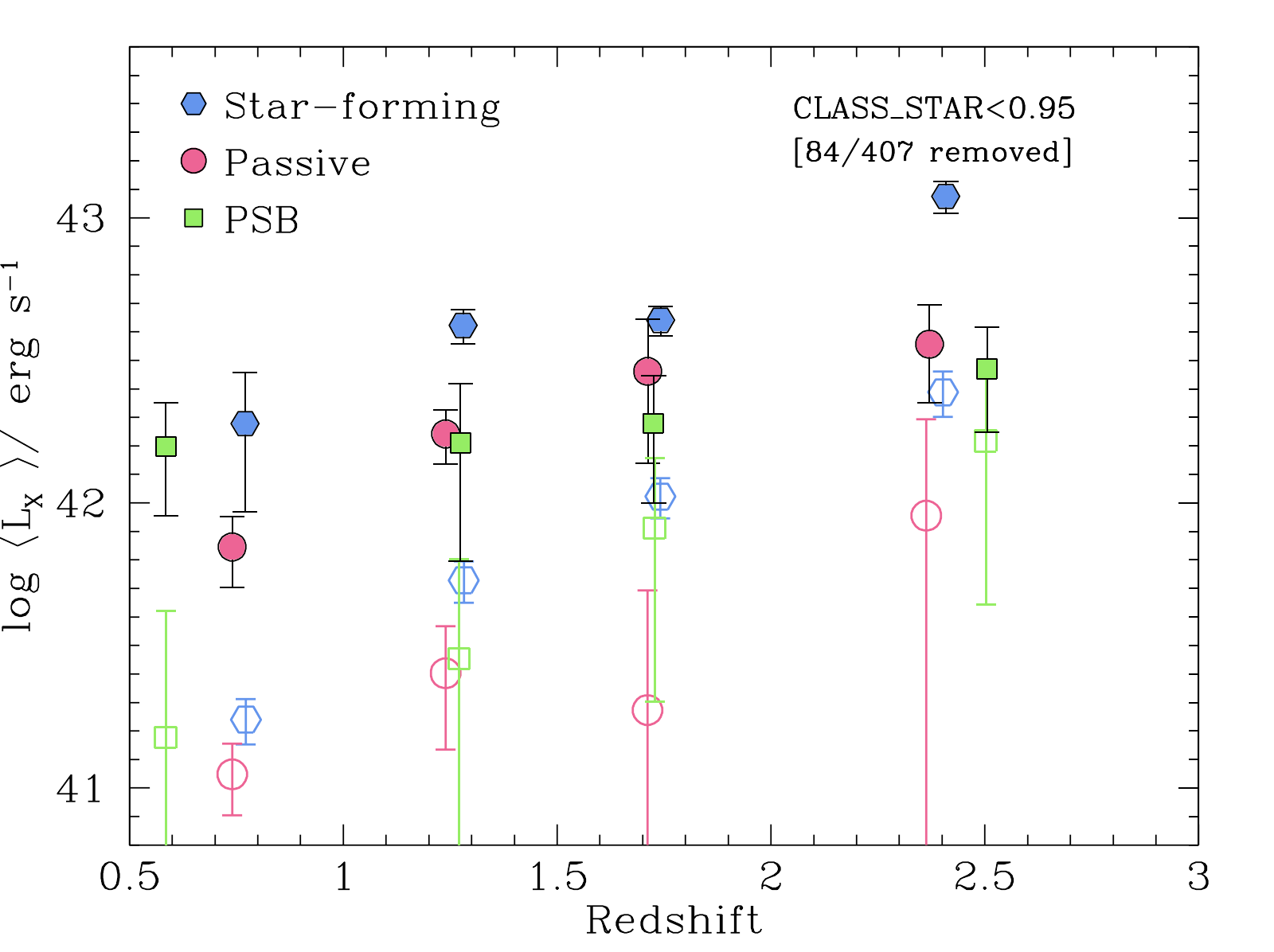}
\includegraphics[width=\columnwidth]{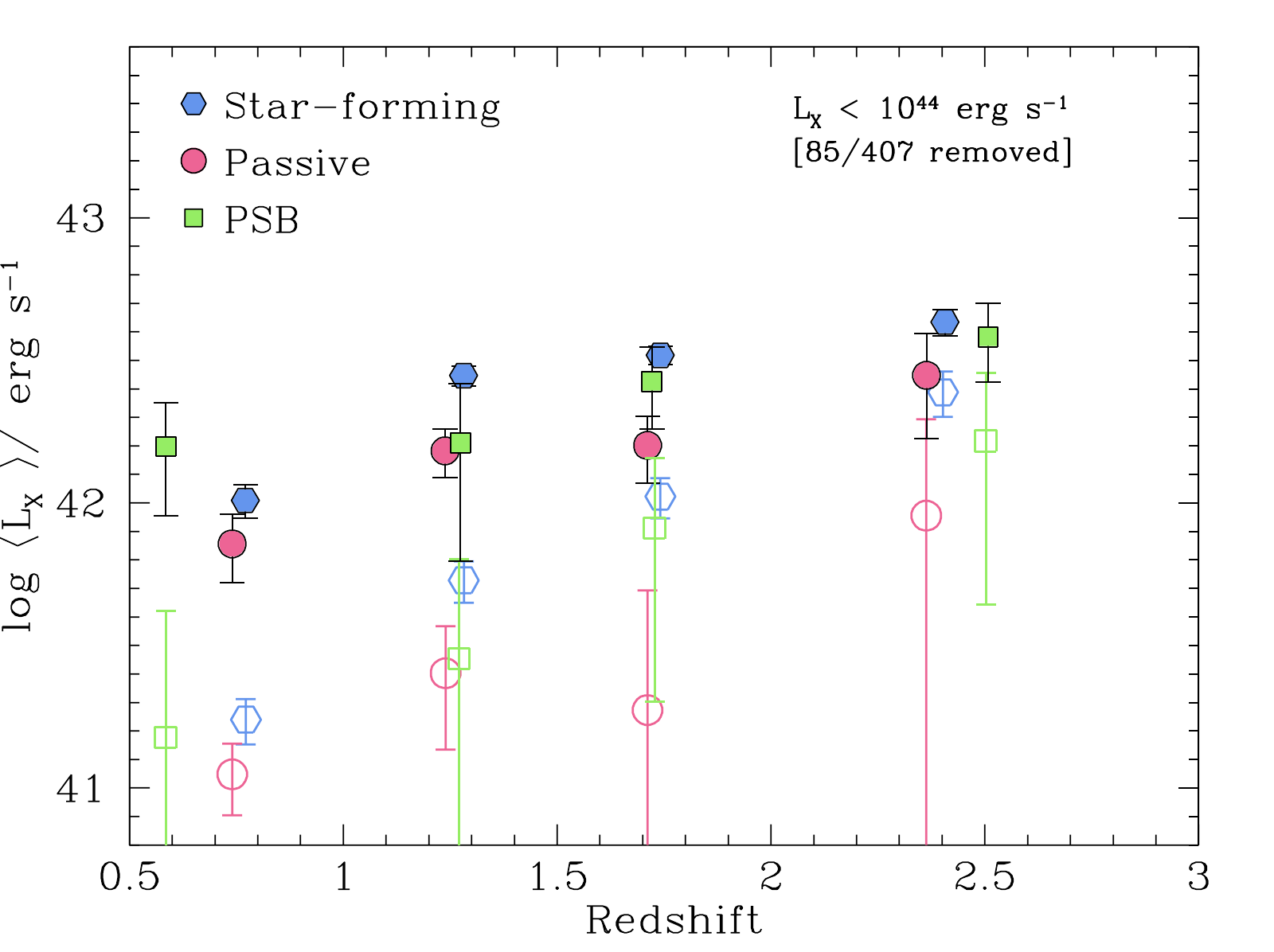} 
 \caption{This figure presents three alternative versions of Figure \ref{fig:lumstack}, to investigate the impact of the criteria for removing quasar-like objects, for which the stellar masses and galaxy classifications may be unreliable. The upper figure shows the full sample, with no quasar candidates removed. The middle plot shows the impact of removing all objects with relatively point-like profiles in our $K$-band
 imaging, according to the 
 \texttt{CLASS\_STAR} parameter. The lower plot shows the impact of removing all objects with X-ray luminosities
  $L_{\rm X}> 10^{44}$ erg s$^{-1}$ (0.5-8 keV).
  }
 \label{fig:A1}
\end{figure}

In Figure \ref{fig:A1} we present three alternative versions 
of
Figure \ref{fig:lumstack}, each showing the average X-ray luminosity as a function of redshift. The three alternatives have no quasar cuts, cuts to exclude all point-like galaxies, and cuts to remove all objects 
with $L_{\rm X}> 10^{44}$ erg s$^{-1}$, respectively. We find that strict luminosity cuts have the biggest impact on the star-forming population, which contains a higher proportion of the most luminous AGN, but overall our key conclusions are unaffected. The PSBs continue to show no excess of AGN activity, instead 
showing average X-ray luminosities that are intermediate between the star-forming and passive galaxies.

\begin{figure}
  \includegraphics[width=\columnwidth]{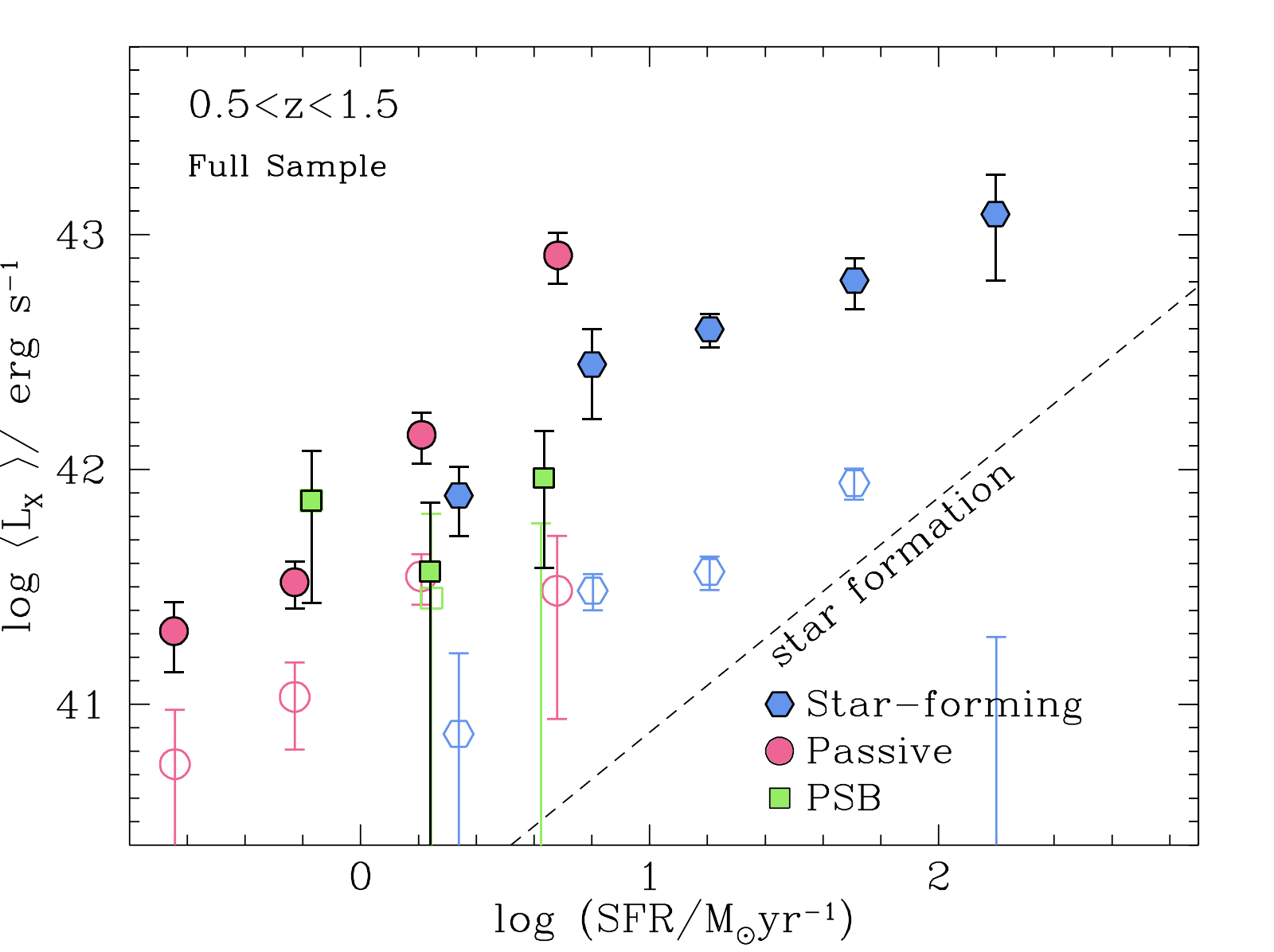}
  \includegraphics[width=\columnwidth]{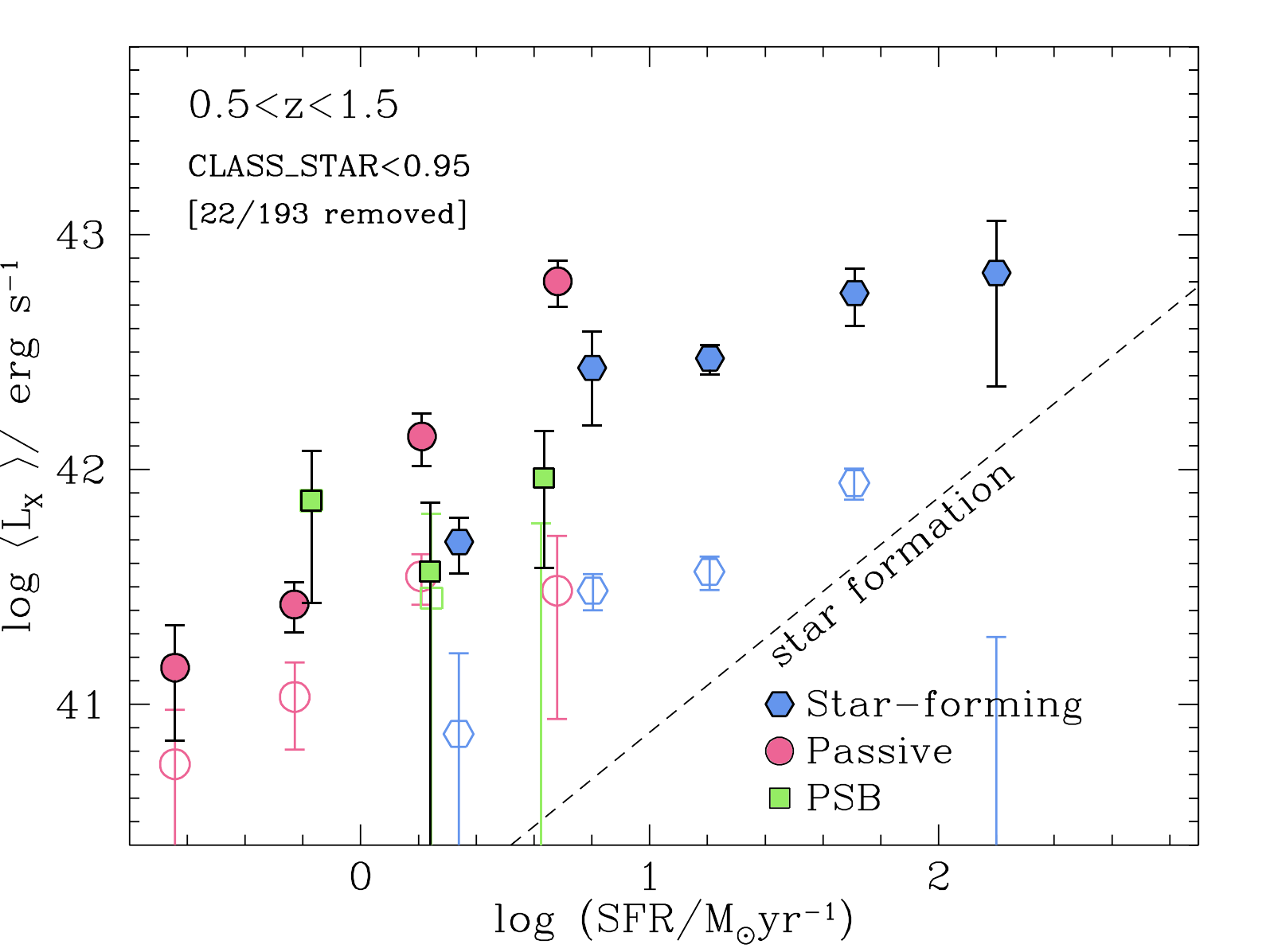}
  \includegraphics[width=\columnwidth]{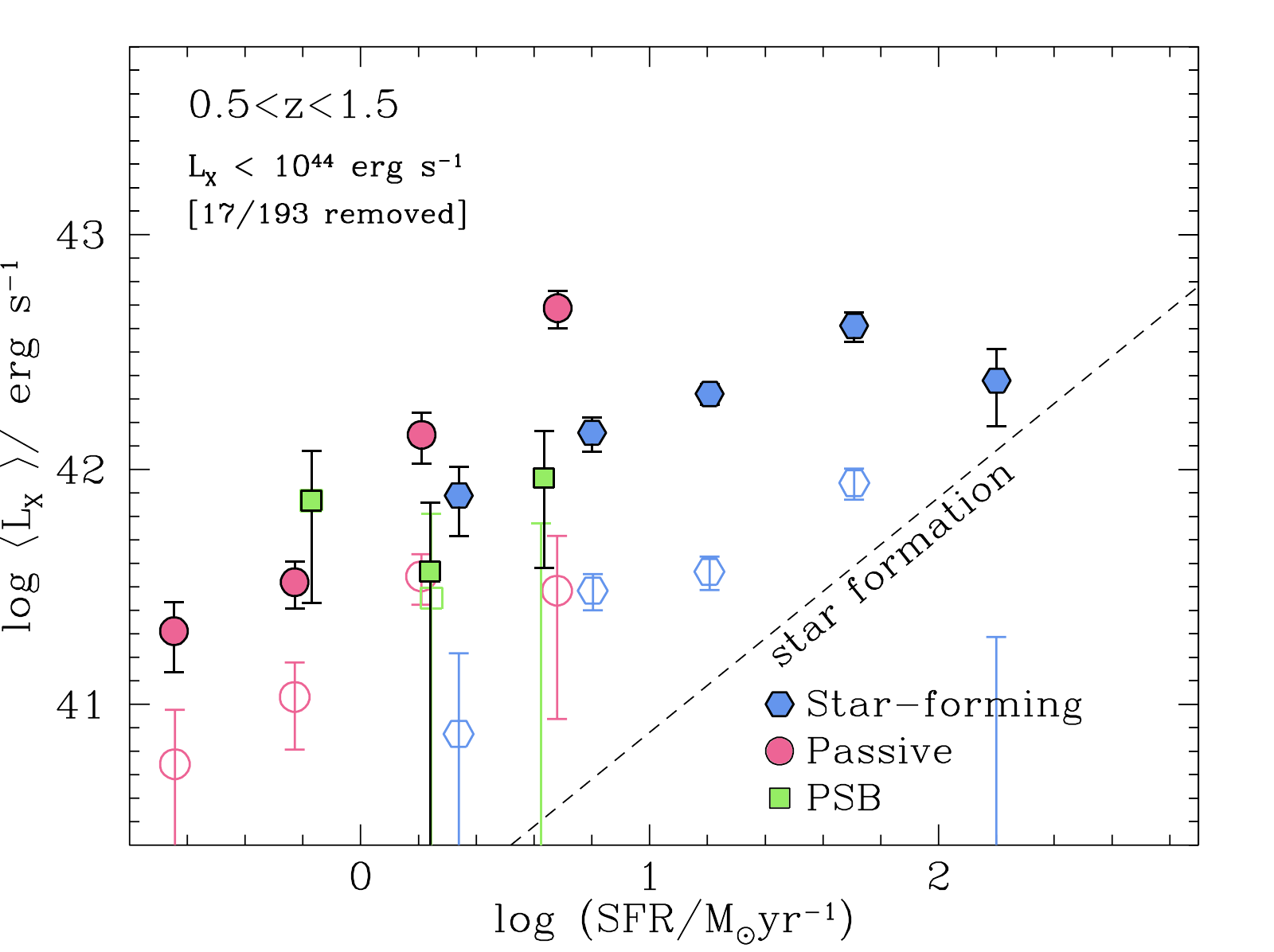}
 \caption{
This figure presents three alternative versions of Figure \ref{fig:sfr} ($0.5<z<1.5$), to investigate the impact of the criteria for removing quasar-like objects, for which the stellar masses and galaxy classifications may be unreliable. The upper figure shows the full sample, with no quasar candidates removed. The middle plot shows the impact of removing all objects with relatively point-like profiles in our $K$-band
 imaging, according to the 
 \texttt{CLASS\_STAR} parameter. The lower plot shows the impact of removing all objects with X-ray luminosities
  $L_{\rm X}> 10^{44}$ erg s$^{-1}$ (0.5-8 keV).
 }
 \label{fig:A2}
\end{figure}

\begin{figure}
  \includegraphics[width=\columnwidth]{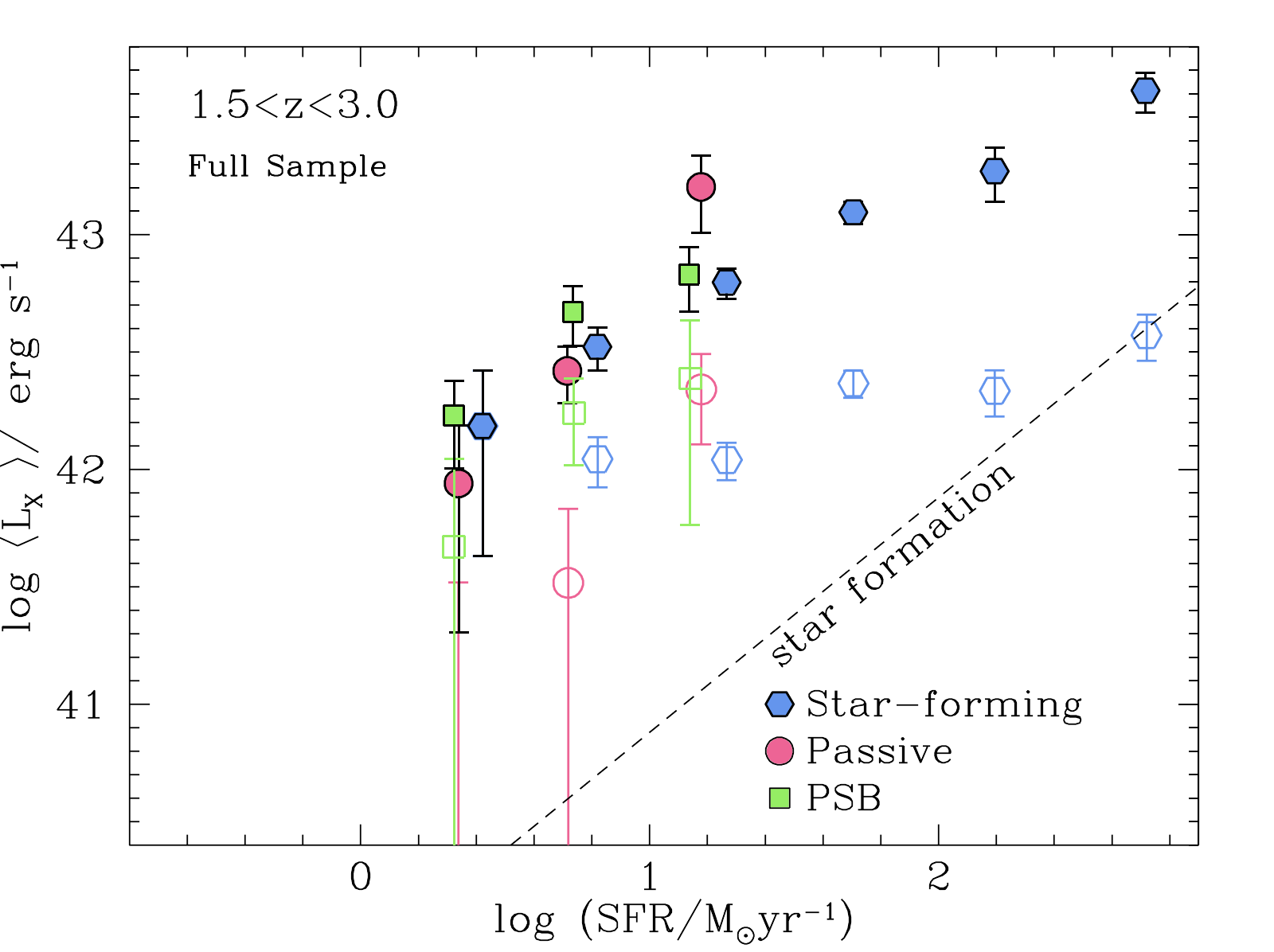}
  \includegraphics[width=\columnwidth]{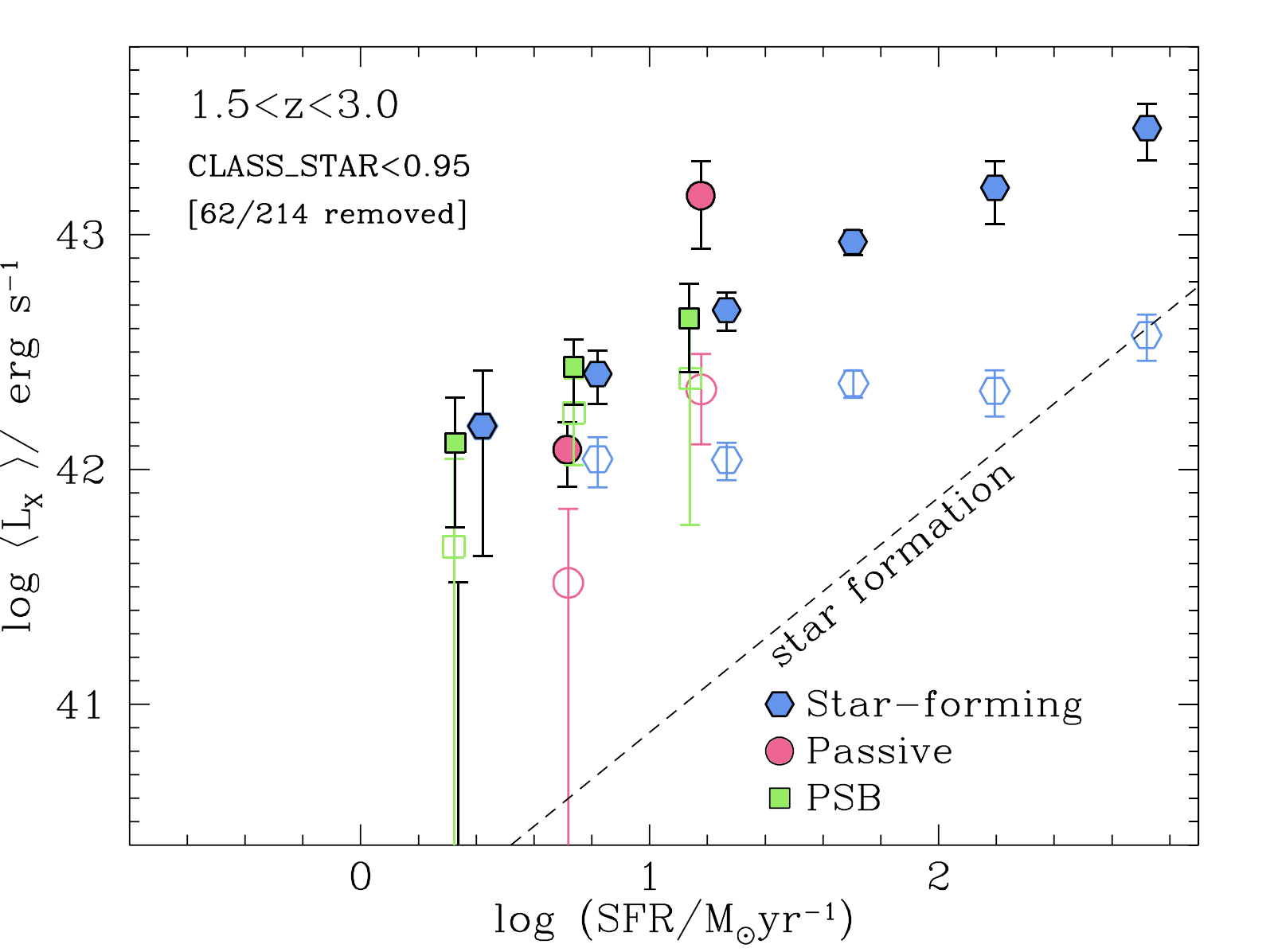}
   \includegraphics[width=\columnwidth]{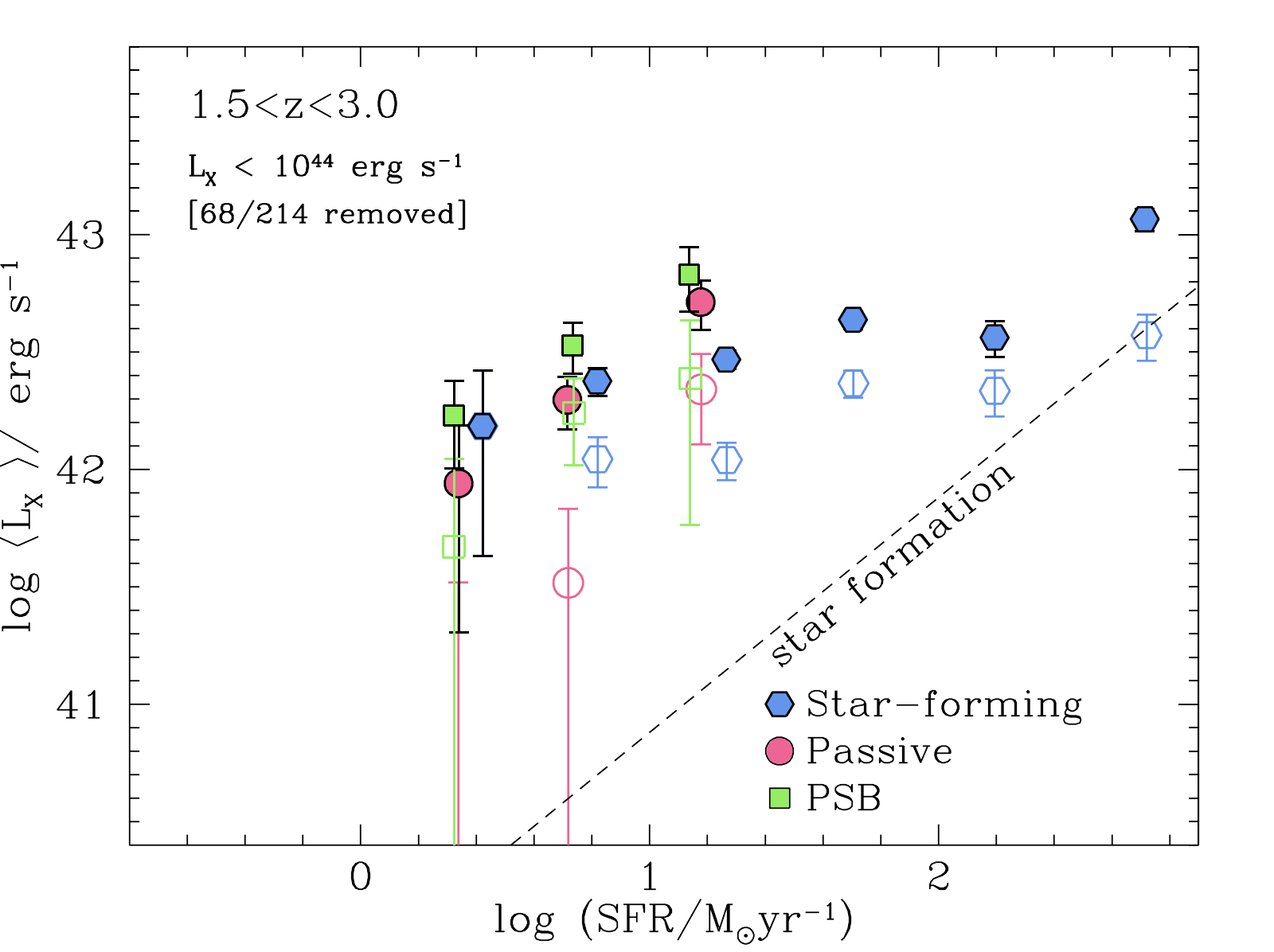}
\caption{
This figure presents three alternative versions of Figure \ref{fig:sfr} ($1.5<z<3.0$), to investigate the impact of the criteria for removing quasar-like objects, for which the stellar masses and galaxy classifications may be unreliable. The upper figure shows the full sample, with no quasar candidates removed. The middle plot shows the impact of removing all objects with relatively point-like profiles in our $K$-band
 imaging, according to the 
 \texttt{CLASS\_STAR} parameter. The lower plot shows the impact of removing all objects with X-ray luminosities
  $L_{\rm X}> 10^{44}$ erg s$^{-1}$ (0.5-8 keV).
}
 \label{fig:A3}
\end{figure}

In Figure \ref{fig:A2} and Figure \ref{fig:A3} we show alternatives
to 
Figure \ref{fig:sfr}, exploring the average X-ray luminosity as a function of SFR, in two broad redshift ranges. The upper two diagrams show only very marginal differences compared to the originals. 
 The strict luminosity cuts inevitably have a major impact in reducing the average X-ray luminosity in the bins at high SFR,  thereby significantly flattening the relation.
 However, we find that overall the evidence for a positive correlation between the mean X-ray luminosity, $\langle L_X \rangle$, and SFR remains for all galaxy populations. At higher redshift (Figure \ref{fig:A3}),  there is marginal evidence for an enhancement in the mean luminosity of PSBs relative to passive and star-forming galaxies after the most luminous X-ray sources are removed, but this is not statistically significant. This trend is  not seen at lower redshift (Figure \ref{fig:A2}).

Overall, we conclude that the precise cuts used to exclude quasar-like objects do not have a major influence on our primary findings.

\bsp	
\label{lastpage}
\end{document}